\documentclass[11pt]{article}

\usepackage{fancyhdr}
\usepackage{fullpage}
\usepackage{amsmath}
\usepackage{amsbsy}
\usepackage{amssymb}
\usepackage{amscd}
\usepackage{amsfonts}
\usepackage{graphics}
\usepackage{verbatim}
\usepackage{epsfig}
\usepackage{xspace}
\usepackage{euscript}
\usepackage{alltt}
\usepackage{float}
\usepackage[colorlinks]{hyperref}
\usepackage{color}
\usepackage{t1enc}
\usepackage{times,exscale}
\usepackage{graphicx,calc}
\usepackage{subfig}
\usepackage{epstopdf}
\usepackage{verbatim}

\usepackage{geometry}
\def\bR{{\mathbf{R}}}
\def\bx{{\mathbf{x}}}

\def\R{{\mathbb{R}}}

\DeclareMathOperator{\Tr}{Tr}

\title{{Spectral quadrature for the first principles study of crystal defects: Application to magnesium}}
\date{\vspace{-10ex}}

\begin{document}
\maketitle
\begin{center}

{Swarnava Ghosh$^{a}$ and Kaushik Bhattacharya$^b$}\\
{$^a$ National Center for Computational Sciences, Oak Ridge National Laboratory, Oak Ridge, TN 37830} \\
{$^b$ Division of Engineering and Applied Science, California Institute of Technology, Pasadena, CA 91125} \\
\date{\today}
\end{center}

\begin{abstract}

We present an accurate and efficient finite-difference formulation and parallel implementation of Kohn-Sham Density (Operator) Functional Theory (DFT) for non periodic systems embedded in a bulk environment. Specifically, employing non-local pseudopotentials, local reformulation of electrostatics, and truncation of the spatial Kohn-Sham Hamiltonian, and the Linear Scaling Spectral Quadrature method to solve for the pointwise electronic fields in real-space and the non-local component of the atomic force, we develop a parallel finite difference framework suitable for distributed memory computing architectures to simulate non-periodic systems embedded in a bulk environment. Choosing examples from magnesium-aluminum alloys, we first demonstrate the convergence of energies and forces with respect to spectral quadrature polynomial order, and the width of the spatially truncated Hamiltonian. Next, we demonstrate the parallel scaling of our framework, and show that the computation time and memory scale linearly with respect to the number of atoms.  Next, we use the developed framework to simulate isolated point defects and their interactions in magnesium-aluminum alloys. Our findings conclude that the binding energies of divacancies, Al solute-vacancy and two Al solute atoms are anisotropic and are dependent on cell size. Furthermore, the binding is favorable in all three cases. 

\end{abstract}

\section{Introduction} \label{Section:Introduction}

Crystal defects, even when present in small concentrations are crucial in determining macroscopic properties of materials. Vacancies, present in parts per million are fundamental to creep, spall and radiation aging. Dislocations, whose density is as small as $10^{-8}$ per atomic row, are the primary carriers of plasticity in metals, solutes present in parts per hundred are responsible for strengthening by interacting with the motion of dislocations, further solutes can aggregate to nucleate a precipitate.

First principles calculations based on Kohn-Sham density functional theory (DFT)\cite{Hohenberg:1964,Kohn:1965} have become central to computational materials research, thereby providing fundamental insights into materials properties and behavior. The success of DFT can be attributed to its excellent predictive power with low accuracy to cost ratio compared to other wavefunction based electronic structure methods \cite{Burke:2012,Becke:2014,Szabo:2012}.  
In spite of its success and widespread use, the efficient solution of Kohn-Sham equations is still computationally daunting, thereby restricting the range of physical systems that can be investigated.  In particular, crystal defects have been particularly challenging because they lead to long-range interactions, the reason why they influence mechanical behavior at small concentrations.

These challenges have led to the development of a number of linear-scaling DFT methods.  However, many of them assume exponential decay of the off-diagonal components of the density matrix $\gamma$, and truncate them to a finite width.  However, while this is reasonable for insulators (since it requires the existence of a band-gap), questions about accuracy remains in the case of metals.  An alternative linear-scaling approach -- the linear scaling spectral Gauss quadratures (LSSGQ) -- was introduced by Suryanarayana, Bhattacharya and Ortiz \cite{Suryanarayana:2012}.  The key idea is to write the density matrix as a (Reimann-Stieltjes) integral over the spectrum (energy states) of the linearized Hamiltonian and then to approximate it using quadratures.   In particular, Gauss quadratures allows one to use the Lanczos algorithm to evaluate the diagonal components of the density matrix at $O(1)$ effort at each spatial point leading to a linear scaling algorithm when one has local pseudopotentials.  Further, the local aspect allows one to introduce variable resolution where fine resolution is maintained where necessary and sub-grid sampling is used in regions of uniform deformation \cite{Suryanarayana:2012,Ponga:2016,Ponga:2020}.  However, it is computationally difficult to compute off-diagonal components of the density matrix using Gauss quadratures, and this makes LSSGQ difficult to extend to nonlocal pseudopotentials.  

Suryanarayana \cite{Suryanarayana:2013} subsequently showed that LSSGQ may be considered a generalization of the Fermi operator expansion (FOE) method using Lagrange polynomials.  This work also proves error estimates for FOE using a polynomial basis, and showed that Gauss quadratures were the most efficient.  It also shows that purification methods using Chebyshev polynomials is equivalent to the use of Clenshaw-Curtis quadratures.  This understanding has led to a series of efficient algorithms using spectral quadratures \cite{Pratapa:2016,Suryanarayana:2018} and applications to first-principles molecular dynamics \cite{Suryanarayana:2018,Zhang:2019,Sharma:2020}.

The first goal of this work is to retain the efficiency of LSSGQ, but also extend it to nonlocal pseudopotentials.  We do so by using LSSGQ for computing the electronic states with atomic positions and Clenshaw-Curtis quadratures for computing the nonlocal contribution of the forces on the atom.  We also introduce a domain decomposition that enables parallel implementation.

The second goal of this work is to use this algorithm to study defects in magnesium.  Magnesium is abundantly available on the earth's crust, the lightest among all commonly used structural metals,  and has among the highest strength to weight ratio \cite{Kainer:2000,Nie:2012,Polmear:1994}.  Aluminum is a commonly used alloying element, and the relative strength of AZ class of magnesium alloys can be attributed to the hexagonal closed-packed (HCP) structure of the magnesium matrix and the $\beta$ phase  Mg$_{17}$Al$_{12}$ precipitates with body-centred cubic structure (space group $I\bar{4}3m$)  \cite{Nie:2012,Kainer:2000,Polmear:1994,Nembach:1997}.

We study isolated vacancy and isolated substitutional aluminum solute in a magnesium lattice along with defect pairs -- vacany pairs, solute-vacancy and solute pairs.
These pairs play an important role in determining the mechanical behavior and processing of magnesium and its alloys.  Vacancy clusters can give rise to prismatic dislocation loops\cite{Ponga:2020,Gavini:2007} or serve as nuclei for voids which in turn are important for spall \cite{Meyers:1983,Ghosh:2019,Gavini:2009}. Such clusters can only form if vacancies in fact can bind. Similarly, aluminum has limited solubility in magnesium and the resulting Mg$_{17}$Al$_{12}$ precipitates play a critical role in strengthening magnesium alloys \cite{Polmear:1994,Kainer:2000,Nembach:1997,Nie:2003,Ghosh:2020}.  This in turn requires both the diffusion of aluminum in a magnesium lattice and an accumulation of aluminum.  The diffusion is greatly aided by the formation of solute-vacancy pairs while the accumulation of aluminum is aided by the binding of aluminum solutes.  Finally, vacancy diffusion is important for dislocation climb, a critical mechanism in creep, and the formation of solute-vacancy pairs are again important.  Previous DFT studies have shown a solute vacancy binding energy that is significantly smaller than that experimentally observed, and non-binding of Al solute pairs raising questions about the mechanism of formation of Al-rich precipitates.   We show that the study of these defects require large computational cells of the type afforded by our algorithm, and suggests that previous contradictory results may have been artifacts of small computational cells.

We introduce our method in Section \ref{Section:Formulation} and describe the numerical implementation in Section \ref{Section:NumericalImplementation}.  We study the convergence and performance of our implementation in Section \ref{Section:Convergence}.  We study defects in Section \ref{Section:Defects}, and close with brief comments in Section \ref{Section:Conclusion}.


\section{Methodology} \label{Section:Formulation}
  
\subsection{Density Operator Formulation of Kohn Sham Density Functional Theory} \label{subsection:DFT}
We consider a cuboidal domain $\Omega$ with $N$ atoms and $N_e$ valence electrons. Let $\bR = \{\bR_1, \bR_2, \ldots, \bR_N \}$ be the positions of the nuclei with valence charges $\{Z_1, Z_2, \ldots, Z_N\}$, respectively. The free energy of this system in Kohn-Sham Density Functional Theory (DFT) \cite{Hohenberg:1964,Kohn:1965} and expressed in terms of density operator \cite{Parr:1989,xbbo_arma_16} is
\begin{equation} \label{Eqn:Energy}
\mathcal{F} (\gamma,\bR) =  2 \Tr\left(-\frac{1}{2}\nabla^2\gamma\right) + E_{xc}(\rho) + 2 \Tr\left(\mathcal{V} \gamma\right) + E_{el}(\rho,\bR) - \theta S(\gamma) \,,
\end{equation}
where $\gamma$ is the density operator, $\Tr(.)$ denotes the trace of an operator, $\rho(\bx)$ := $2 \gamma(\bx,\bx)$ is the electron density. The first term in (\ref{Eqn:Energy}) is the kinetic energy of non-interacting electrons and $E_{xc}$ is the exchange correlation energy in the local density approximation (LDA).  Specifically we use the Perdew-Wang parameterization \cite{Perdew:1992} of the correlation energy calculated by Ceperley-Alder \cite{Ceperley:1980}.
The third term, $2 \Tr\left(\mathcal{V} \gamma\right)$, is the contribution of the non-local pseudopotential to the free energy.     $\mathcal{V}$ is the non-local pseudopotential operator given by 
$$
\mathcal{V}(\bx,\bx') = \sum_{J} \mathcal{V}_J (\bx,\bx') =  \sum_{J} \sum_{lm} c_{Jl} \chi_{Jlm} (\bx,,\bR_J)  \chi_{Jlm} (\bx',\bR_J) 
$$
where $\mathcal{V}_{J}$ is the contribution to the non-local pseudopotential operator from the $J^{th}$ atom, and the summation over $J$ is over all nuclei whose supports of the non-local projectors $\chi_{Jlm}$ overlap with domain $\Omega$.
The fourth term is the electrostatic energy which is expressed 
within the local reformulation framework \cite{Suryanarayana:2010,Pask:2005,Ghosh:2016} as
\begin{equation} \label{Eqn:ElecEnergyReformulation}
E_{el}(\rho,\bR) = \sup_{\phi}  \bigg \{ - \frac{1}{8 \pi} \int_{\Omega} |\nabla \phi(\bx,\bR)|^2 \, \mathrm{d\bx} + \int_{\Omega}(\rho(\bx)+ b(\bx,\bR)) \phi(\bx,\bR) \, \mathrm{d\bx} \bigg \} + E_{self}(\bR)\,\,, 
\end{equation}
where $\phi$ denotes the electrostatic potential, $b$ represents the total pseudocharge density of the nuclei and $E_{self}(\bR)$ is the self energy of the nuclei
\begin{equation}\label{Eqn:Eself}
E_{self}(\bR) = -\frac{1}{2} \sum_{I} \int_{\Omega} b_I (\bx,\bR) V_I(\bx,\bR) \mathrm{d\bx} \,\,,
\end{equation}
where $b_I$ is the pseudocharge density of the $I^{th}$ nucleus generating the potential $V_I$ (the summation over $I$ is over all nuclei in $\R^3$ whose pseudocharge densities overlap with $\Omega$).  The final term in (\ref{Eqn:Energy}) is the electronic entropy arising from the partial occupancies of the electronic states at a finite electronic temperature $\theta$ with 
\begin{eqnarray}
S(\gamma) = -2 k_{B} \Tr \big[ (\gamma \log \gamma  + (\mathcal{I} - \gamma) \log (\mathcal{I} - \gamma)\big] 
\end{eqnarray}
and $\mathcal{I}$ the identity operator.

The \emph{ground state} in DFT obtained by minimizing the functional $\mathcal{F} (\gamma,\bR)$ over all atomic positions $\bR$ and all density operators $\gamma$ associated with $N_e$ electrons.  It is convenient to nest this minimization problem by first calculating the \emph{electronic ground state}:
\begin{eqnarray}\label{Eq:EGS}
\hat{\mathcal{F}}(\bR) = \inf_{\{\gamma \ \text{s.t.}  \ 2\Tr(\gamma) = N_e \}} \ \ \mathcal{F} (\gamma,\bR) 
\end{eqnarray}
and then relaxing over all atomic configurations
\begin{equation}\label{Eq:GS}
\mathcal{F}_0 = \inf_{\bR} \hat{\mathcal{F}}(\bR) .
\end{equation}
 The Euler-Lagrange equation to the variational problem (\ref{Eq:EGS})  is a nonlinear fixed-point problem:
\begin{equation}\label{Eqn:FixedPoint}
\gamma = g(\mathcal{H},\lambda_f;\theta) = \left(1+\exp \left(\frac{\mathcal{H}-\lambda_f \mathcal{I}}{k_{B}\theta} \right)\right)^{-1}
\end{equation}
where $k_{B}\theta$ is the electronic smearing, the Fermi energy $\lambda_f$ is the Lagrange multiplier employed to enforce the constraint on the number of electrons, and the Hamiltonian $\mathcal{H}$ is 
\begin{equation}
\mathcal{H} = -\frac{1}{2} \nabla^2 + V_{xc} + \phi + \mathcal{V} \,\,,
\end{equation}
with $V_{xc} = \delta E_{xc}/\delta \rho$  the exchange-correlation potential and $\phi$  the solution of the Poisson equation
\begin{equation}\label{Eqn:Poisson}
-\frac{1}{4 \pi} \nabla^2 \phi (\bx,\bR) = \rho(\bx) + b(\bx,\bR) 
\end{equation}
subject to appropriate boundary conditions.  Note that  $V_{eff} = V_{xc} + \phi$ is local (diagonal operator) and hence its action on a function $f$ is given by $(V_{eff} f)(\bx) = V_{eff}(\bx)f(\bx)$. The action of the non-local pseudopotential operator $\mathcal{V}$ is given by
\begin{eqnarray}
(\mathcal{V}f)(\bx) = \sum_{J} \mathcal{V}_{J}f = \sum_{J} \sum_{lm} c_{Jl} \chi_{Jlm} (\bx) \int_{\Omega} \chi_{Jlm} (\bx',\bR_J) f(\bx') \mathrm{d\bx'} .
\end{eqnarray}

After evaluating the electronic ground state, the free energy is calculated using the functional:
\begin{eqnarray}\label{Eqn:FreeEnergy}
\mathcal{\hat{F}}(\bR) &=& U+ E_{xc}(\rho) - \int_{\Omega} V_{xc} (\rho) \rho(\bx) \mathrm{d\bx} + \frac{1}{2} \int_{\Omega} (b(\bx,\bR)-\rho(\bx)) \phi(\bx,\bR) \mathrm{d\bx} \nonumber \\ 
&&  - \theta S 
+ E_{self}(\bR)\,\,,
\end{eqnarray}
where $U = 2\Tr(\gamma \mathcal{H})$ is the \emph{band structure energy}. 
The atomic force on the $J^{th}$ atom is calculated using the expression
\begin{eqnarray}\label{Eq:AtomicForce}
\mathbf{f}_J &=& \frac{\partial \mathcal{\hat{F}}(\bR)}{\partial \bR_J} = 
\int_{\Omega} \nabla b_J(\bx,\bR_J)\phi(\bx,\bR) \mathrm{d\bx} - 4 \Tr \left(\mathcal{V}_J \nabla \gamma \right) 
\end{eqnarray}
where the first term is  local (recall that $\phi(\bx,\bR)$ depends only on the local or diagonal components $\rho(\bx) = \gamma(\bx,\bx)$ of the density operator) while the second term is non-local and requires the off-diagonal terms of the density operator.


\subsection{Linear Scaling Spectral Quadrature} \label{subsection:SQ}

We follow the Spectral Quadrature (SQ) method \cite{Ponga:2016,Ponga:2020,Suryanarayana:2012,Suryanarayana:2013,Pratapa:2016,Suryanarayana:2018} for solving the DFT problem. The key idea is to ground state quantities as (Reimann-Stieltjes) integrals over the spectrum of the Hamiltonian.  Given the Fermi level $\lambda_F$ and the Hamiltonian $\mathcal{H}$, we can use (\ref{Eqn:FixedPoint}) to write
\begin{equation}\label{Eqn:SpectralTheorem}
(\eta,\gamma \eta) = \int_{\sigma} g(\mathcal{H},\lambda_f;\theta) \mathrm{d}\mu_{\eta,\eta}(\lambda) = \int_{\sigma} g(\lambda,\lambda_f;\theta) \mathrm{d}\mu_{\eta,\eta}(\lambda) 
\end{equation}
for any function $\eta$  where $\sigma$ is the spectrum of $\mathcal{H}$, and $\mu_{\eta,\eta}$ is the spectral measure of $\mathcal{H}$ contracted with $\eta$.  We use the spectral theorem \cite{Rudin:1973} to obtain the second equality.    We can now use quadratures to approximate the integral.  In this work we use Gauss quadratures to find the electronic ground state and Clenshaw-Curtis quadratures to find the force on an atom.

\paragraph*{Spectral Gauss quadrature} We follow the linear scaling spectral Gauss quadrature (LSSGQ) method of Suryanarayana, Bhattacharya and Ortiz \cite{Suryanarayana:2012} that exploits the structure of Gauss quadratures to evaluate the electronic ground state.  In Gauss quadratures, we approximate any function $f(\lambda)$ in terms of Lagrange polynomials $P_k^{\eta}(\lambda)$ as
\begin{equation}
 f(\lambda) \approx \sum_{k=0}^K f(\lambda_k^{\eta}) P_k^{\eta}(\lambda)\,\,
\end{equation}
where $K$ is the degree of the expansion and $\lambda_k^{\eta}$ are the spectral nodes. We can use this expansion to approximate the integral of the function $f$ over the spectrum of $\mathcal{H}$,
\begin{equation}\label{Eqn:SpectralIntegralApprox}
I[f] =  \int_{\sigma} f({\lambda}) \, \mathrm{d}\mu_{\eta,\eta}({\lambda}) \approx \sum_{k=0}^K f(\lambda_k^{\eta}) \left( \int_{\sigma}  P_k^{\eta}(\lambda) \mathrm{d}\mu_{\eta,\eta}(\lambda) \right) = \sum_{k=1}^{K} {w}_k^{\eta} f({\lambda}_k^{\eta}) 
\end{equation}
where the spectral weight ${w}_k^{\eta}$ denotes the integral $\int_{\sigma}  P_k^{\eta}(\lambda) \mathrm{d}\mu_{\eta,\eta}(\lambda)$. 

Now, consider a discretization of the computational domain using a regular finite difference grid with $N_d$ points, and let $\{ \eta_q\}_{q=1}^{N_d}$ be a set of orthonormal basis functions associated with this discretization such that $\eta_q$ is compactly supported near the $q^{th}$ grid point.  We can then approximate the integrals that make up the electronic ground state (cf. (\ref{Eqn:FreeEnergy})) as 
\begin{eqnarray}\label{Eqn:RSIntegrals}
U = \sum_{q=1}^{N_d} u^q, \quad, S = \sum_{q=1}^{N_d} s^q, \quad N_e =  \sum_{q=1}^{N_d} \rho^q,
\end{eqnarray}
where
\begin{eqnarray}
u^q &=& 2 \int_{\sigma} \lambda \, g({\lambda},\lambda_f;\theta) \, \mathrm{d}\mu_{\eta_q,\eta_q}({\lambda})  \,  
\approx  \sum_{k=1}^{K} {w}_k^{\eta_q} \lambda_k^{\eta_q} g(\lambda_k^{\eta_q},\lambda_f;\theta) ,  \label{eq:uq}
  \\
s^q &=& -2 k_{B} \int_{\sigma} s({\lambda},\lambda_f;\theta) \, \mathrm{d}\mu_{\eta_q,\eta_q}({\lambda})  \, 
 \approx  -2 k_{B}  \sum_{k=1}^{K} {w}_k^{\eta_q} s(\lambda_k^{\eta_q},\lambda_f;\theta) , \label{eq:sq}
\\
\rho^q &=& 2  \int_{\sigma} g({\lambda},\lambda_f;\theta) \, \mathrm{d}\mu_{\eta_q,\eta_q}({\lambda})  \,  
\approx  2  \sum_{k=1}^{K} {w}_k^{\eta_q} g(\lambda_k^{\eta_q},\lambda_f;\theta) ,  \label{eq:rhoq}
\end{eqnarray}
 $g(\lambda,\lambda_f;\theta)$ is the Fermi-Dirac function
\begin{equation}\label{Eqn:FermiDirac}
g(\lambda,\lambda_f;\theta) = \frac{1}{1+\exp(\frac{\lambda-\lambda_f}{k_{B}\theta})} \,\,,
\end{equation}
and $s(\lambda,\lambda_f;\theta)$ is the pointwise contribution to the electronic entropy
\begin{equation}
s(\lambda,\lambda_f;\theta) = g(\lambda,\lambda_f;\theta) \log g(\lambda,\lambda_f;\theta) + (1-g(\lambda,\lambda_f;\theta))\log(1-g(\lambda,\lambda_f;\theta)).
\end{equation}

In Gauss quadrature, the weights $\{{w}_k^{\eta}\}_{k=1}^K$ are fixed apriori, and the spectral nodes $\{ \lambda_k^{\eta} \}_{k=1}^K$ are treated as unknowns. An efficient way of evaluating the spectral weights and nodes at any grid point $q$ is by employing the Lanczos iteration
\begin{eqnarray}\label{Eqn:Lanczos}
b_{k+1}^q v_{k+1}^q &=& (\mathcal{H}-a_{k+1}^q)v_{k}^q - b_k^q v_{k-1}^q\,,\,\, k=0,1,\ldots, K-1 \nonumber \\
v_{-1}^q &=& 0\,\,, v_0^q = \eta_q\,\,, b_0^q = 1\,\,,
\end{eqnarray}
where $a_{k+1}^q = (\mathcal{H}v_k^q, v_k^q)$, $k=0,1,\ldots,K-1$, and $b_k^q$ is computed such that $||v_k^q||=1,k=0,1,\ldots,K-1$. We denote the Jacobi matrix $\hat{J}_K^q$ as
\begin{equation}
\hat{J}_K^q = \begin{pmatrix}
a_1^q & b_1^q \\
b_1^q & a_2^q & b_2^q \\
    & \ddots & \ddots & \ddots \\
    &  & b_{K-2}^q & a_{K-1}^q & b_{K-1}^q \\
    &   &        & b_{K-1}^q & a_K^q \\
\end{pmatrix}
\end{equation}
The nodes $\{w_k^{\eta_q}\}_{k=1}^K$ are calculated as the eigenvalues of $\hat{J}_K^q$, and the weights $\{\lambda_k^{\eta_q}\}_{k=1}^K$ are the squares of the first elements of the normalized eigenvectors of $\hat{J}_K^q$.

The key observations of LSSGQ are that (i) the number of quadrature points $K$ is independent of system size and (ii) the vectors $v_k^q$ remains zero except for a small region around the grid point $q$ during Lanczos iteration (\ref{Eqn:Lanczos}).  Therefore, the evaluation of the spectral nodes at each grid point is $O(1)$ as is the evaluation of all the electronic quantities of interest (cf. (\ref{Eqn:FreeEnergy}) and (\ref{Eqn:RSIntegrals})).  This enables us to evaluate the electronic ground state at linear cost.

We further observe that this limited support of the vectors $v_k^q$ enables the restriction of the Hamiltonian $\mathcal{H}$ to an appropriate subspace of the real-space computational domain $\Omega$. This enables us to use domain decomposition in our numerical implementation discussed in Section. \ref{Section:NumericalImplementation}.

\paragraph*{Spectral Clenshaw-Curtis quadrature} 

While the Gauss quadrature provides a very efficient approach to evaluating  the electronic ground state since it depends only on the diagonal terms of the density matrix, it is unable to evaluate quantities like the contribution of the non-local pseudopotential to the atomic force that depends on the off-diagonal components.  Therefore, we use the spectral Clenshaw-Curtis quadrature \cite{Suryanarayana:2013,Pratapa:2016,Suryanarayana:2018}.

The Hamiltonian is first scaled and shifted such that its spectrum lies in the interval $[-1, 1]$ 
\begin{equation}
\mathcal{\hat{H}} = \frac{\mathcal{H}-\mathcal{\omega}\mathcal{I}}{\mathcal{\tau}} \,\,
\end{equation} 
where $\omega= (\lambda^{max} + \lambda^{min})/2$, $\tau=(\lambda^{max} - \lambda^{min})/2$, and $\lambda^{max}$, $\lambda^{min}$ are the maximum and minimum eigenvalues of $\mathcal{H}$, respectively.
Given any function $\eta$, we can rewrite (\ref{Eqn:SpectralTheorem}) using the scaled and shifted Hamiltonian $\mathcal{\hat{H}}$:
\begin{equation}
(\eta, \gamma \eta) = \int_{\sigma} g(\mathcal{H},\lambda_f;\theta) \mathrm{d}\mu_{\eta,\eta}(\lambda) = \int_{-1}^{1} g(\mathcal{\hat{H}},\hat{\lambda}_f;\hat{\theta}) \mathrm{d}\mu_{\eta,\eta}(\hat{\lambda}) = \int_{-1}^{1} g(\hat{\lambda},\hat{\lambda}_f;\hat{\theta}) \mathrm{d}\mu_{\eta,\eta}(\hat{\lambda})
\end{equation}
where $\hat{\lambda_f} = (\lambda_f-\omega)/\tau$ and $\hat{\theta}= \theta/\tau$ are the scaled Fermi energy and the scaled electronic temperature respectively. In the Clenshaw-Curtis variant of the spectral quadrature, any function $f(\hat{\lambda})$ is expanded in terms of Chebyshev polynomials $T_k^{\eta}(\hat{\lambda})$ as:
\begin{equation}
 f(\hat{\lambda}) \approx \sum_{k=0}^K f(\hat{\lambda}_k^{\eta}) T_k^{\eta}(\hat{\lambda})\,\,
\end{equation}
where $K$ is the degree of the expansion, $\hat{\lambda}_k^{\eta}$ are the spectral nodes which are fixed in Clenshaw-Curtis quadrature. 
The  non-local component of the atomic force \cite{Pratapa:2016,Suryanarayana:2018} as
\begin{eqnarray}\label{Eqn:NLforceSQ}
4\Tr \left(\mathcal{V}_J \nabla \gamma \right) &\approx& 4 \sum_{p=1}^{N_d} \left( \eta_q, (\mathcal{V}_J \nabla g(\mathcal{H},\lambda_f;\theta))\eta_q  \right)
= 4 \sum_{q=1}^{N_d} \left( \eta_q, (\mathcal{V}_J \nabla g(\mathcal{\hat{H}},\hat{\lambda}_f;\hat{\theta}))\eta_q  \right)  \nonumber \\
&\approx& 4 \sum_{q=1}^{N_d} \sum_{k=0}^K c_k^q \eta_q^* \mathcal{V}_J \nabla v_k^q \,\,
\end{eqnarray}
where $c_k^q$ are constants
\begin{equation}\label{Eqn:ChebyshevWeights}
c_k^q = \frac{2}{\pi} \int_{-1}^{1} \frac{g(\lambda,\hat{\lambda}_f;\hat{\theta})T_k(\lambda)}{\sqrt{1-\lambda^2}} \,\, \mathrm{d}\lambda \,\,,
\end{equation}
and $v_{k+1}^q=T_k(\mathcal{\hat{H}})\eta_q$ are functions which are determined by the three term recurrence relation for Chebyshev polynomials:
\begin{eqnarray}\label{Eqn:Chebyshev}
v_{k+1}^q &=& 2 \mathcal{\hat{H}} v_k^q - v_{k-1}^q \\ \nonumber
v_1^q &=& \mathcal{\hat{H}} \eta_q, v_0^q = \eta_q \,\,.
\end{eqnarray}
Once again, the number of quadratures is independent of the system size and therefore this evaluation scales linearly.
%
%


\section{Numerical Implementation} \label{Section:NumericalImplementation}
\begin{figure} \centering
{\includegraphics[keepaspectratio=true,width=1\textwidth]{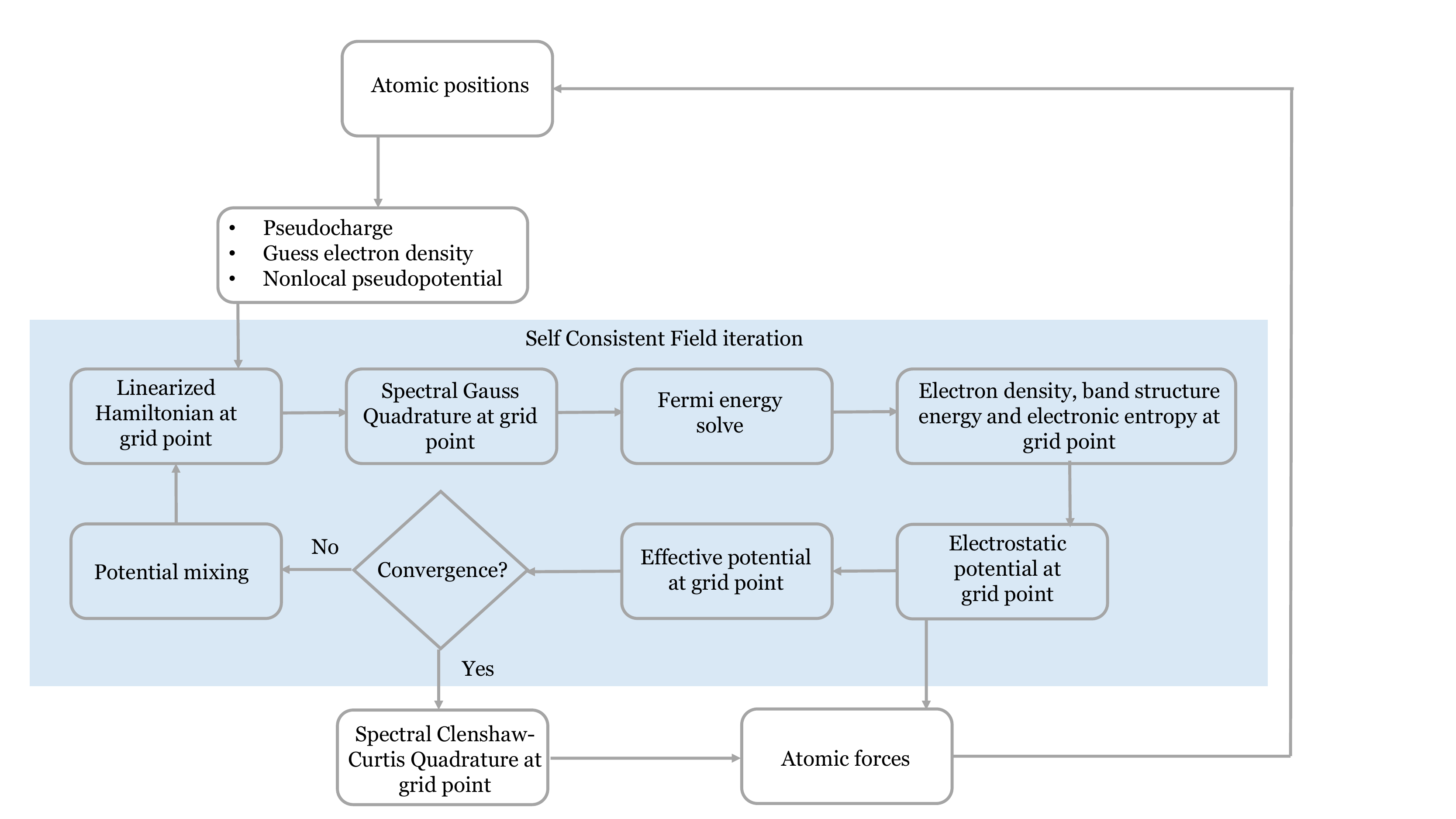}}
\caption{Flowchart of the Self Consistent Field iteration for solving the DFT problem.} \label{Fig:Flowchart}
\end{figure}

Fig. \ref{Fig:Flowchart} shows the flowchart of the scheme employed to solve the DFT problem. The non-linear fixed point problem (Eqn. \ref{Eqn:FixedPoint}) is solved using the self-Consistent field (SCF) iteration.  Briefly, given a charge density and electrostatic potential, we construct a linearized Hamiltonian which is used to compute the spectral weights nodes using Lanczos iteration from which the updated electronic states including an updated charge density and electrostatic potential are determined; the process iterates till convergence.  Once the electronic ground state is established, the atomic relaxation for the overall ground state (Eqn. \ref{Eq:GS}) is achieved using the Non-Linear Conjugate Gradient (NLCG) method \cite{Shewchuk:1994}. 

We discuss further details of the key components of the implementation below.  In this paper, we are interested in isolated defects.  Therefore, we consider a computational domain that is embedded in an infinite perfect crystal.  The method can be extended to other situations including periodic boundary conditions and isolated clusters surrounded by vacuum.

\paragraph*{Discretization} 
Let $\Omega = L_1 \times L_2 \times L_3$ be a cuboidal computational domain aligned with the coordinate axis, and let it be discretized using a regular 
$n_1 \times n_2 \times n_3$ grid so that the grid spacing is $h_i$ in the $i^{th}$ coordinate where $L_i = n_i h_i$.  We index the grid points with $q$ and let $K_\Omega$ be the set of all grid points.  We discretize the gradient and Laplace operators using finite differences on this grid \cite{Ghosh:2017a,Ghosh:2017b}.

\begin{figure} \centering
\subfloat[region of influence]{\includegraphics[keepaspectratio=true,width=0.6\textwidth]{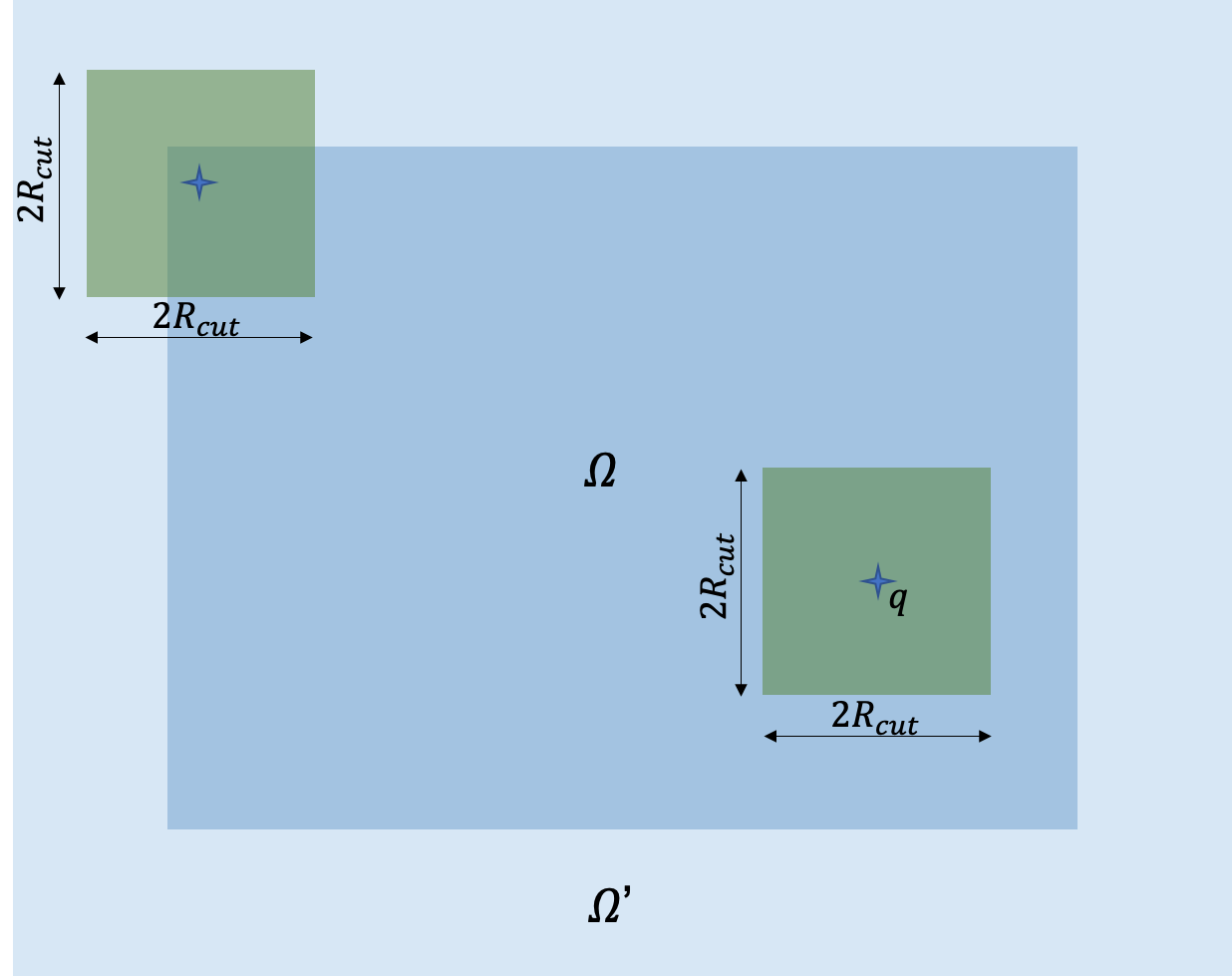}\label{Fig:Hdomain}}
\subfloat[neighbor communicator]{\includegraphics[keepaspectratio=true,width=0.5\textwidth]{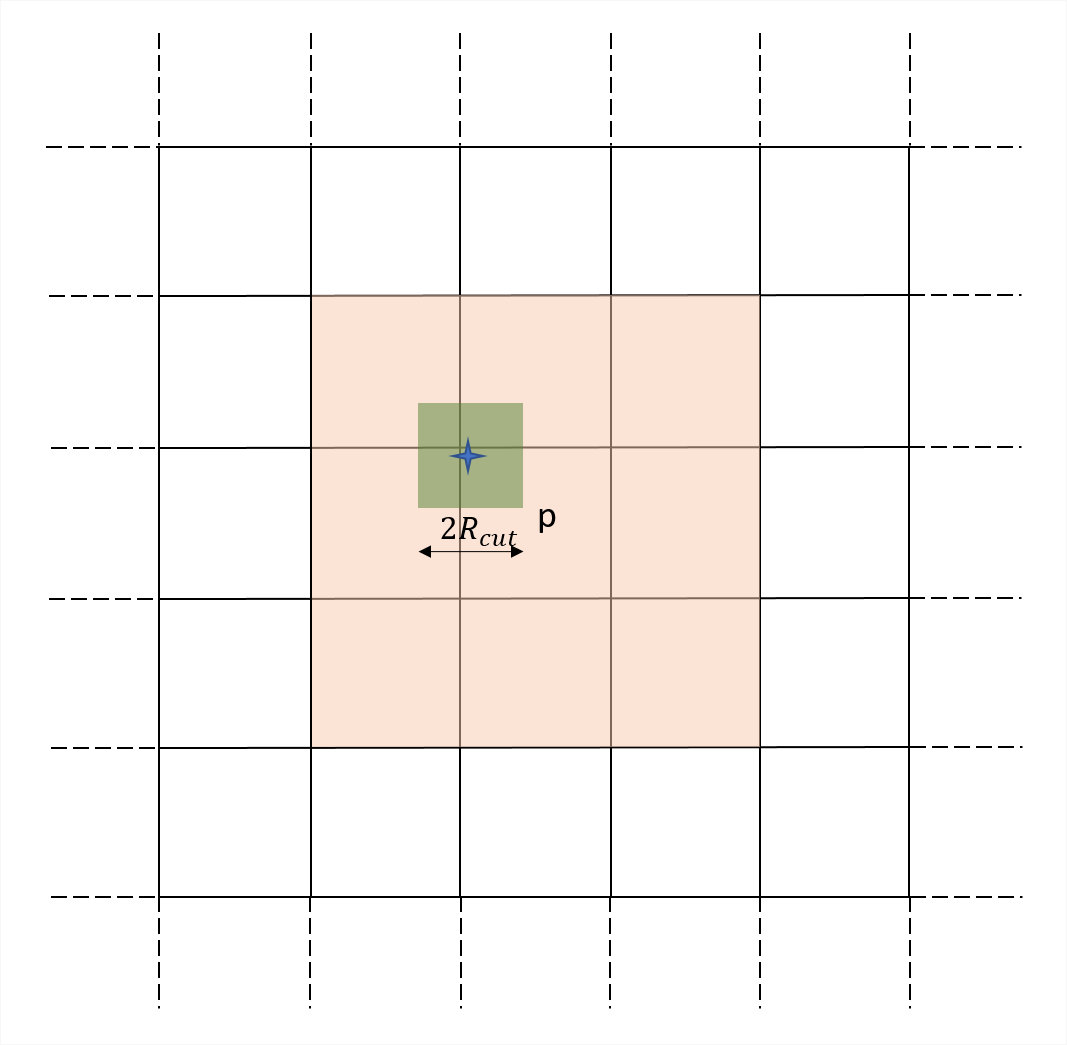}\label{Fig:Communicator}}
\caption{(a) Region of influence associated with the $q^{th}$ grid point and extended domain. (b) Neighbor communicator in domain decomposition. The orange region is the union of partitioned domains that influence the process $p$.}
\end{figure}

\paragraph*{Hamiltonian at each grid point}
In each iteration of SCF, we need to determine the spectral nodes and weights at each grid point. As evident from (\ref{Eqn:Lanczos}) and (\ref{Eqn:Chebyshev}), this requires the calculation of the action of Hamiltonian on vectors $v_k^q$. These vectors are compactly supported around a ball centered at the $q^{th}$ grid point.   Therefore, it is sufficient to work with the Hamiltonian $\mathcal{H}^q$ projected on to a smaller subspace around the $q^{th}$ grid point.  Specifically, we work with a cuboidal region of side $2R_{cut}$ and centered at the $q^{th}$ grid point which we call \emph{region of influence}.   This is shown schematically in Fig. \ref{Fig:Hdomain}.  The controllable parameter $R_{cut}$ depends on the order of the quadrature which in turn depends on material properties and electronic smearing temperature $\theta$ \cite{Pratapa:2016,Suryanarayana:2018}.

%
%

For grid points close to the edge of $\Omega$, the nodal Hamiltonian can extend beyond the computational domain for grid points which are near the boundary of $\Omega$. 
However, since we consider problems where our computational domain is embedded in an infinite crystal, we have to compute the Hamiltonian on an extended region $\Omega'$ of size $(L_1+2R_{cut})\times(L_2+2R_{cut})\times(L_3+2R_{cut})$, and centered at the origin.  The values of the Hamiltonian associated with the annular $\Omega' \setminus \Omega$ is obtained using precomputed electronic fields ($\rho$ and $\phi$) for the perfect crystal.

\paragraph*{Domain decomposition}

We use domain decomposition for parallel implementation.  The computational domain is partitioned into disjoint domains, $\Omega = \bigcup\limits_{p=1}^{N_p} \Omega_p$, where $\Omega_p$ denotes the domain local to the $p^{th}$ processor, and $N_p$ is the total number of processors. The collection of grid points belonging to the $p^{th}$ processor is denoted by $K_{\Omega}^p$, where  $K_{\Omega} = \bigcup\limits_{p=1}^{N_p} K_{\Omega}^p$, and $K_{\Omega}^p\bigcap  K_{\Omega}^q = \emptyset$ (null set) for $p \neq q$. In our implementation, we use the DMDA class available through the Portable, Extensible Toolkit for Scientific Computation (PETSc) \cite{Petsc1,Petsc2} for mesh management. The communication between processes is handled by Message Passing Interface (MPI) libraries \cite{Gropp:1999,Gropp:2014}.

The region of influence of an grid point $q$ owned by a process $p$ (i.e. $q\in  K_{\Omega}^p$) may extend to the spatial regions owned by neighboring processes. In such a situation, the values of the effective potential $V_{eff}$ from neighboring processes are communicated to the process $p$ using an MPI communicator. In Fig. \ref{Fig:Communicator} we schematically illustrate the parallel communication pattern involved for communicating the effective potential $V_{eff}$ from neighboring processes. This reduces the number of MPI related calls otherwise required for matrix vector products.

\paragraph*{Spectral weights, nodes and electronic fields}

In each SCF iteration, the spectral weights $\{w_k^q\}_{k=1}^K$ and nodes $\{\lambda_k^q\}_{k=1}^K$ are computed at every grid point $q\in  K_{\Omega}$ from the projected Hamiltonian $\mathcal{H}^q$.  Overall, this is computationally the most expensive part of the method.  However, the computation is local and with no MPI related calls.  Further, we do not explicitly store the matrices, and their multiplication with a vector is performed in a matrix-free manner.   These lead to excellent parallel efficiency.

%
%

Once the spectral weights $\{w_k^q\}_{k=1}^K$ and spectral nodes $\{\lambda_k^q\}_{k=1}^K$ are computed at all the grid points $q\in  K_{\Omega}^p$ for all the processes $p$, we first solve the following equation for the Fermi energy $\lambda_f$:
\begin{equation}\label{Eqn:Ne}
N_e = 2  \sum_{p=1}^{N_p} \sum_{q \in K_{\Omega}^p} \sum_{k=1}^K w_k^q g(\lambda_k^q,\lambda_f;\theta) \,\,.
\end{equation}
We utilize Brent's algorithm \cite{Brent:1971} for this purpose. The summation across the polynomial degree $k$ and the grid point $q$ is performed locally in each processes, and the summation across the processes $p$ is performed with one global MPI communication call. 

Once the Fermi energy is calculated, we calculate the point-wise band structure energy $u^q$, the point-wise entropy $s^q$ and the point-wise electron density $\rho^q$, 
following (\ref{eq:uq}), (\ref{eq:sq}), (\ref{eq:rhoq}).  These are all local.

\paragraph*{Electrostatic and effective potential}
Once the electron density $\rho$ is calculated at the grid points, we calculate the total electrostatic potential $\phi$ at the grid points by solving the Poisson equation (\ref{Eqn:Poisson}) on $\Omega$ subject to Dirichlet boundary conditions obtained from the perfect crystal outside using the generalized minimal residual algorithm (GMRES) \cite{Saad:1986}. Once the electrostatic potential is calculated at every grid point $q \in \Omega$, we calculate the effective potential $V_{eff}$ at every grid point
\begin{equation}
V_{eff}^q = V_{xc}(\rho^q) + \phi^q \,\,,
\end{equation}
where $V_{xc}(\rho^q)$ is the exchange correlation potential. 

The convergence of the SCF iteration is accelerated by employing mixing schemes. In every SCF iteration, we mix the effective potential $V_{eff}$, where we have the option of employing Anderson mixing \cite{Anderson:1965}, Pulay mixing scheme  \cite{Pulay:1980} and its periodic \cite{Banerjee:2016:Mixing} and restarted \cite{Pratapa:2015:Mixing} variants.

We check the convergence of SCF iteration by calculating the normalized error in the effective potential. 

\paragraph*{Free energy}
The free energy is computed once the SCF iteration has converged using a discrete version of Eqn. \ref{Eqn:FreeEnergy}
\begin{eqnarray} \label{Eqn:DiscreteEnergy}
\mathcal{\hat{F}}(\bR) \approx \mathcal{\hat{F}}^h(\bR) = \sum_{p=1}^{N_p} \sum_{q \in K_{\Omega}^p} \left[ u^q + h_1 h_2 h_3 \left\lbrace \left(\varepsilon_{xc}^q- V_{xc}^q\right) \rho^q  + \frac{1}{2}\left(b^q-\rho^q\right)\phi^q \right\rbrace - 2 \theta s^q \right] + E_{self}^h\,\,,
\end{eqnarray} 
where $ \varepsilon_{xc}^q$ is the contribution of the exchange correlation energy at the $q^{th}$ grid point, and $ E_{self}^h$ is the discrete representation of the self energy (\ref{Eqn:Eself}):
\begin{equation}\label{Eqn:DiscreteSelfEnergy}
E_{self}^h = -\frac{1}{2} (h_1h_2h_3) \sum_{p=1}^{N_p} \sum_{q\in K_{\Omega}^p} \sum_I b_I^q V_I^q
\end{equation}
In evaluating them, the sum over the grid points are carried out locally on each MPI process followed by a global sum across all the MPI processes $p$.

\paragraph*{Atomic forces}
The final step is the computation of the atomic forces (Eqn. \ref{Eq:AtomicForce}).  This  has two parts, the contributions of the local pseudopotential and the non-local pseudopotentials. The contribution of the local pseudopotential to the atomic force is calculated as
\begin{equation}\label{Eqn:DiscreteLocalForce}
{\bf{f}}_{J,l} = \int_{\Omega} \nabla b_J(\bx,\bR_J)\phi(\bx,\bR) \mathrm{d\bx} \approx h_1h_2h_3 \sum_{p=1}^{N_p} \sum_{q\in K_{\Omega}^p} (\nabla_{h}b_J )^q \phi^q
\end{equation}
where $\nabla_h$ is the gradient operator in the discrete setting.  The summation over the the grid points $q$ is local to every process, followed by one summation over the MPI processes.  

The  non-local contribution to the atomic force is calculated by employing the Clenshaw-Curtis quadrature  described in Section \ref{subsection:SQ}. At each grid point $q \in K_{\Omega}$, we calculate the discrete Chebyshev vectors ${\bf v}_q^k$ using the iterative scheme (\ref{Eqn:Chebyshev}), and the Clenshaw-Curtis quadrature weights $c_q^k$ using (\ref{Eqn:ChebyshevWeights}) with the discrete nodal Hamiltonian ${\bf H}_q$. The non-local force is given by
\begin{equation}
{\bf{f}}_{J,nl} = - 4 \Tr(\mathcal{V}_J\nabla \gamma) \approx -4 \sum_{p=1}^{N_p}  \sum_{q\in K_{\Omega}^p} {\bf e}_q^T \mathcal{V}_J^q \, \nabla_h \left( \sum_{k=0}^K c_q^k {\bf v}_k^q \right).
\end{equation}
The calculation of the atomic forces scales linearly with the number of atoms. 

\section{Convergence and Performance} \label{Section:Convergence}

We set the electronic temperature to be $k_B \theta$ = 0.03333 Ha, and use Troullier-Martins non-local pseudopotentials \cite{Troullier:1991}.  These yield lattice parameters of $a$ = 6.043 Bohr, $c$ = 9.848 Bohr and $c/a$ ratio of 1.629 for hexagonal closed packed (HCP) magnesium, and $a$=19.58 Bohr for body centered cubic (BCC) Mg$_{17}$Al$_{12}$.  These agree with  the previously reported values of the lattice constants: $a$=5.972 Bohr, $c/a$=1.61 (DFT) \cite{Chou:1986}, $a$=6.066 Bohr, $c/a$=1.623 (experiment)\cite{Kittel,Koster:1961} for Mg, and the equilibrium lattice constant of the Mg$_{17}$Al$_{12}$ or MgAl phase is $a$ = 19.96 Bohr (DFT) \cite{Duan:2011}, and $a$ = 19.653 Bohr (experiment) \cite{Zhang:2005} for Mg$_{17}$Al$_{12}$.

\paragraph{Convergence of spectral quadrature}\label{subsection:Npl}
\begin{figure}\centering
\subfloat[energy]{\includegraphics[keepaspectratio=true,width=0.49\textwidth]{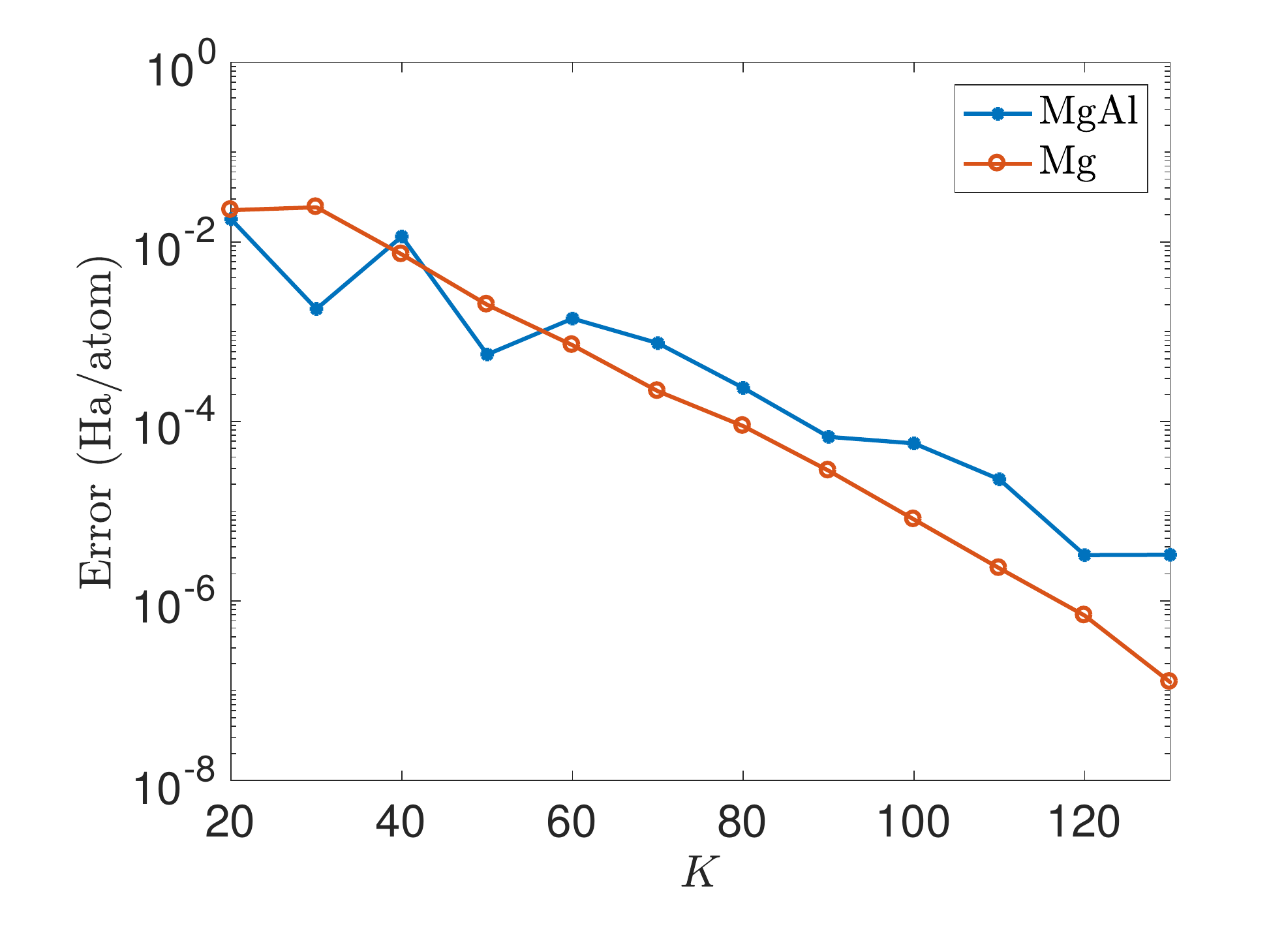}\label{Fig:ConvergenceEnergyPolynomial}} 
\subfloat[force]{\includegraphics[keepaspectratio=true,width=0.49\textwidth]{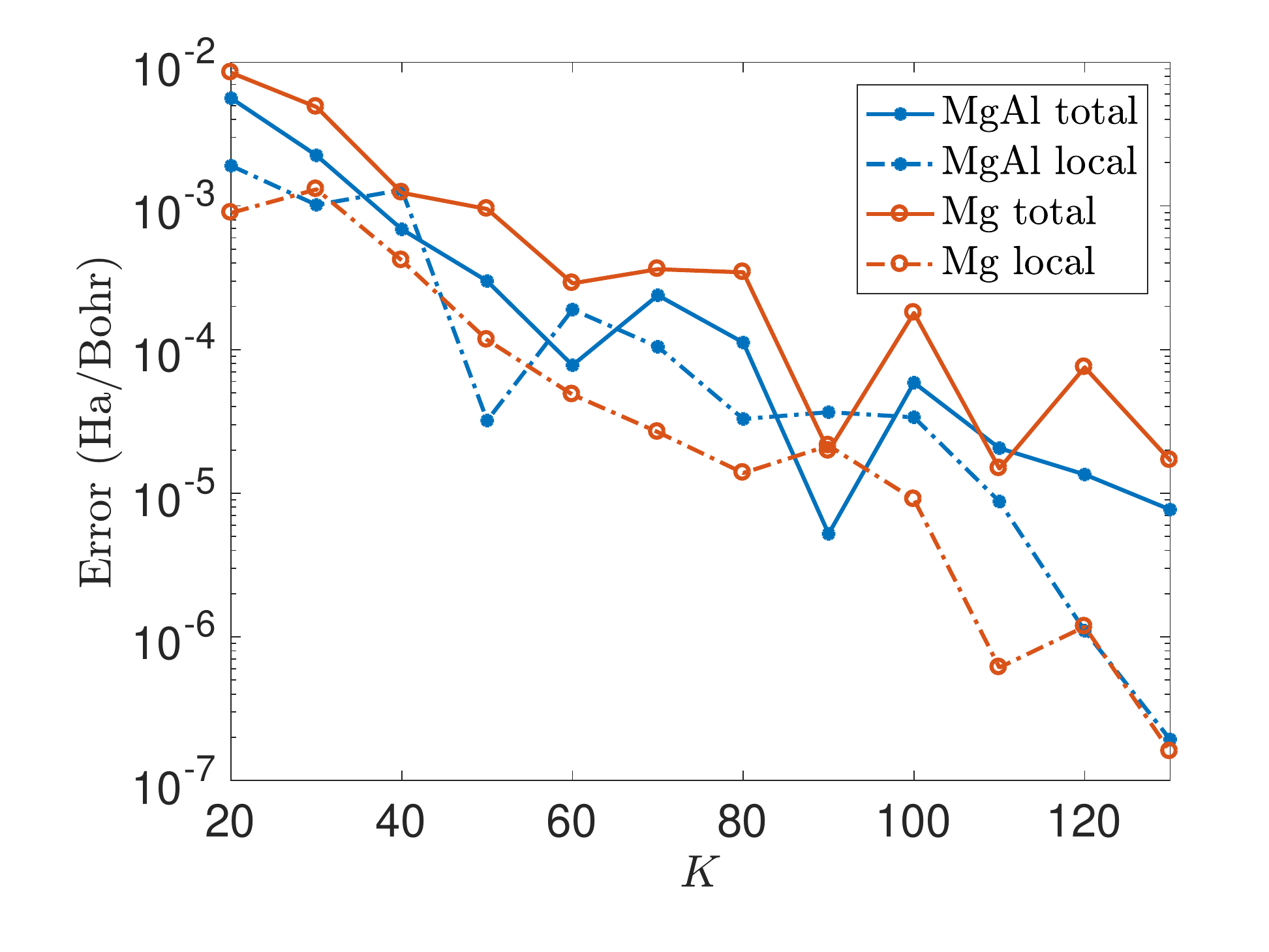}\label{Fig:ConvergenceForcePolynomial}}
\caption{Convergence of (a) energy and (b) force as a function of degree of spectral quadrature.} 
\end{figure}

We first verify the convergence with respect to spectral quadrature.  We take the same degree $K$ for both the Gauss and the Clenshaw-Curtis quadrature.  We study HCP Mg and BCC MgAl.  For each of these systems, we randomly perturb the atoms from the ground state to obtain the test configuration, and use a mesh of $h=0.6$ Bohr for pure Mg and $h=0.5$ Bohr for MgAl.
%

Fig. \ref{Fig:ConvergenceEnergyPolynomial} shows the convergence of the energy and Fig. \ref{Fig:ConvergenceForcePolynomial} shows the convergence of atomic forces with $K$.  The error is computed with respect to the reference that is taken to be the solutions for $K$=$240$.  The decay is not monotone because neither the free energy functional (Eqn. \ref{Eqn:Energy}) nor the atomic forces (Eqn. \ref{Eq:AtomicForce}) is  variational with respect to $K$.  However, it broadly follows an exponential decay (Error  $\approx A e^{-\beta K}$).  The best fit $\beta$ for the various cases is shown in Table \ref{Table:Convergence}.  The pure Mg system has a higher rate of convergence than the MgAl system. Furthermore, in both the cases, the rate of convergence in total energy and the local component of the atomic force is similar, whereas the rate of convergence in the total force, which require calculating the non-local force component is smaller by almost a factor 2. This is so because the local component of the atomic force and energy use only Gauss quadrature which depends on Lagrange polynomials whereas the non-local component of atomic forces depends on Clenshaw Curtis quadrature with Chebyshev polynomials.  The former is known to be more accurate than the latter \cite{Suryanarayana:2013}.

Since we require an accuracy of about $1\times 10^{-3}$ and $1\times 10^{-4}$ Ha/atom in our energy, we need between $60$ and $80$ for Mg, and between $60$ and $90$ for MgAl.  Similarly, since we require an accuracy between $1\times 10^{-3}$ and $1\times 10^{-4}$ Ha/Bohr in the total force, we need between $50$ and $100$ for Mg and MgAl.  These guide our further calculations.

\begin{table}[t]
\caption{Rates of convergence $\beta$ of the total energy, local force and the total force with the spectral quadrature polynomial order $K$.}
\label{Table:Convergence}
\centering
\begin{tabular}{cccc}
\hline
System & Energy & Force (local) & Force (total)   \\
\hline
MgAl &  0.070 & 0.075 & 0.036  \\
Mg &  0.109 &  0.088 & 0.042 \\
\hline
\end{tabular}
\end{table}

\paragraph{Convergence with radius of truncation}\label{subsection:Rcut}
\begin{figure}
\centering
\subfloat[energy]{\includegraphics[keepaspectratio=true,width=0.49\textwidth]{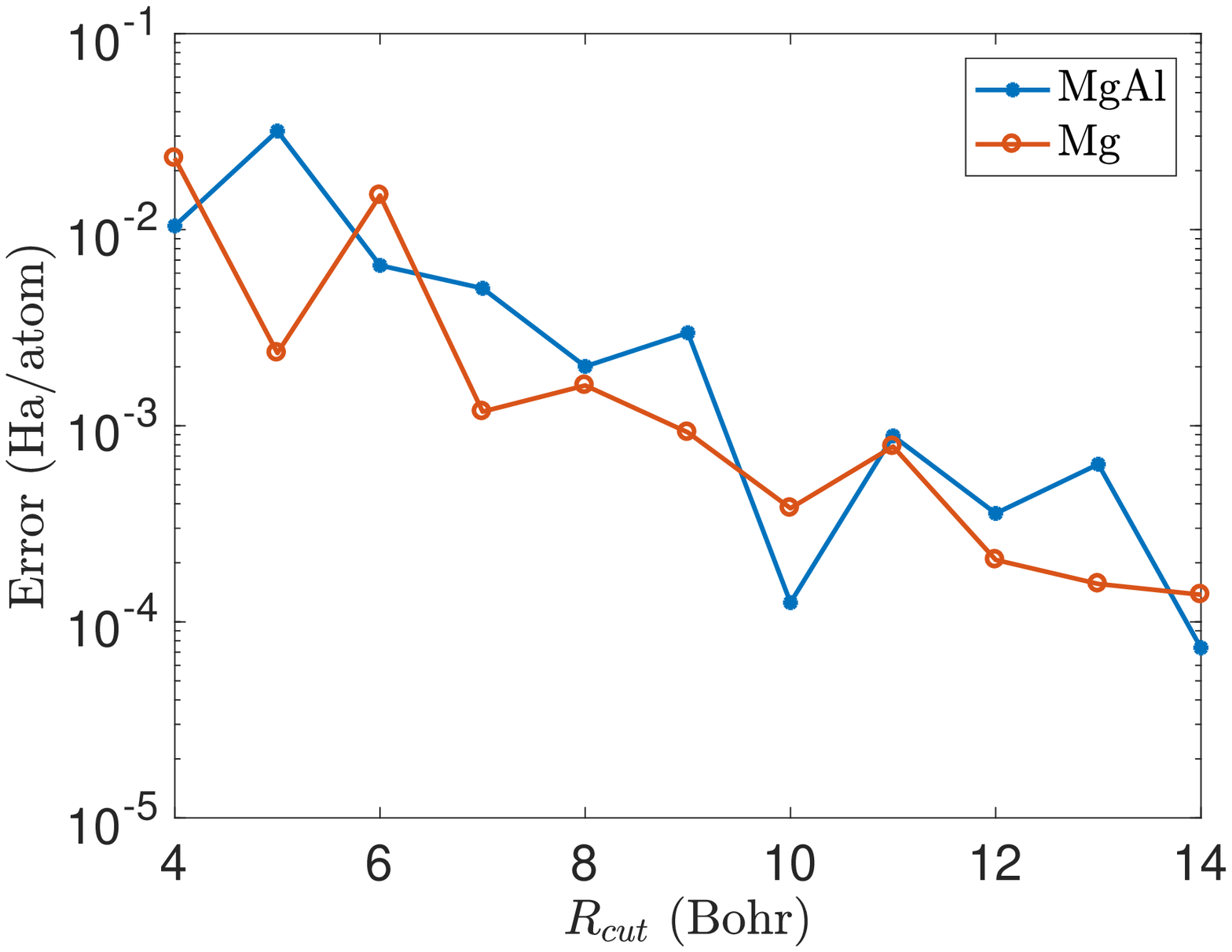}\label{Fig:ConvergenceEnergyRcut}}
\subfloat[force]{\includegraphics[keepaspectratio=true,width=0.49\textwidth]{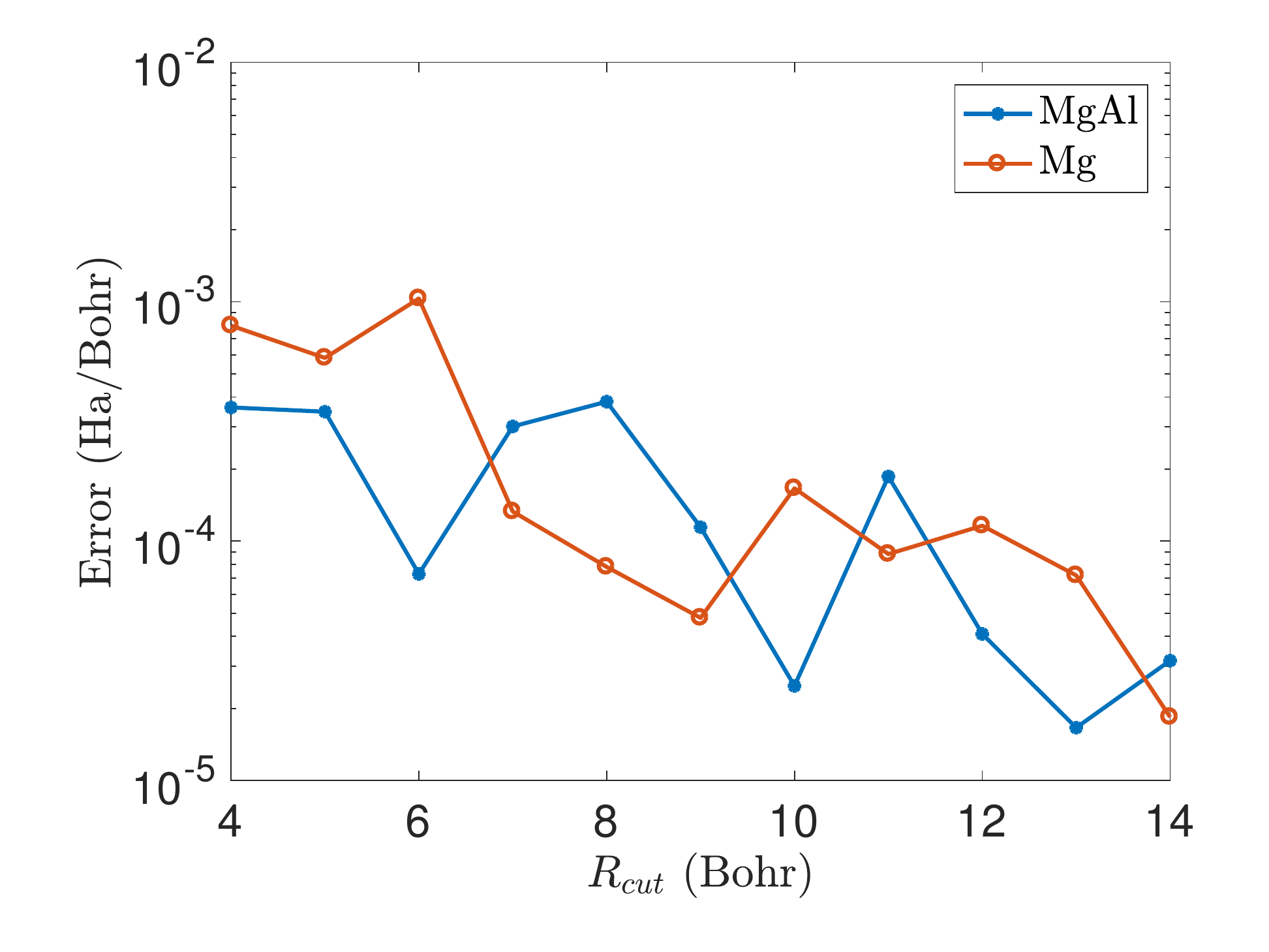}\label{Fig:ConvergenceForceRcut}}
\caption{Convergence of (a) energy and (b) total atomic force as a function of truncation radius  $R_{cut}$ of the nodal Hamiltonian.} 
\end{figure}

A loose upper bound of the size $R$ of Lanczos vectors in (\ref{Eqn:Lanczos}) is given by $2R \leq h n_o K $, where $h$ is the average mesh spacing, and $n_o$ is the finite difference order.   This suggests a choice of $2R_{cut} = h n_o K $.  Using the choice of parameters used for this study, $R_{cut}  \approx 288$ Bohr.  This
 would make the calculations extremely expensive.  However, we now show that it is not necessary by studying the error as a function of $R_{cut}$.

For the Mg and and MgAl systems, we use a Lanczos polynomial degree $K_L$ = $80$ and a Chebyshev polynomial degree $K_C$ = $100$, and vary $R_{cut}$ from $4.0$ to $14.0$ Bohr. Fig. \ref{Fig:ConvergenceEnergyRcut}  and \ref{Fig:ConvergenceForceRcut} shows the error in energy and atomic force as a function of $R_{cut}$. The reference values of energy and atomic force is calculated using an $R_{cut}$ = $16$ Bohr. From these figures, we observe that as $R_{cut}$ increases, the error in energy and forces decrease.  We once again observe generally exponential decay, though it is not monotone (also see \cite{Pratapa:2016}).   This allows us to treat $R_{cut}$ as a controllable approximation parameter.
   
\paragraph{Scaling and performance}\label{subsection:Performance}
      
\begin{figure}[t]
\centering
\subfloat[strong scaling]{\includegraphics[keepaspectratio=true,width=0.49\textwidth]{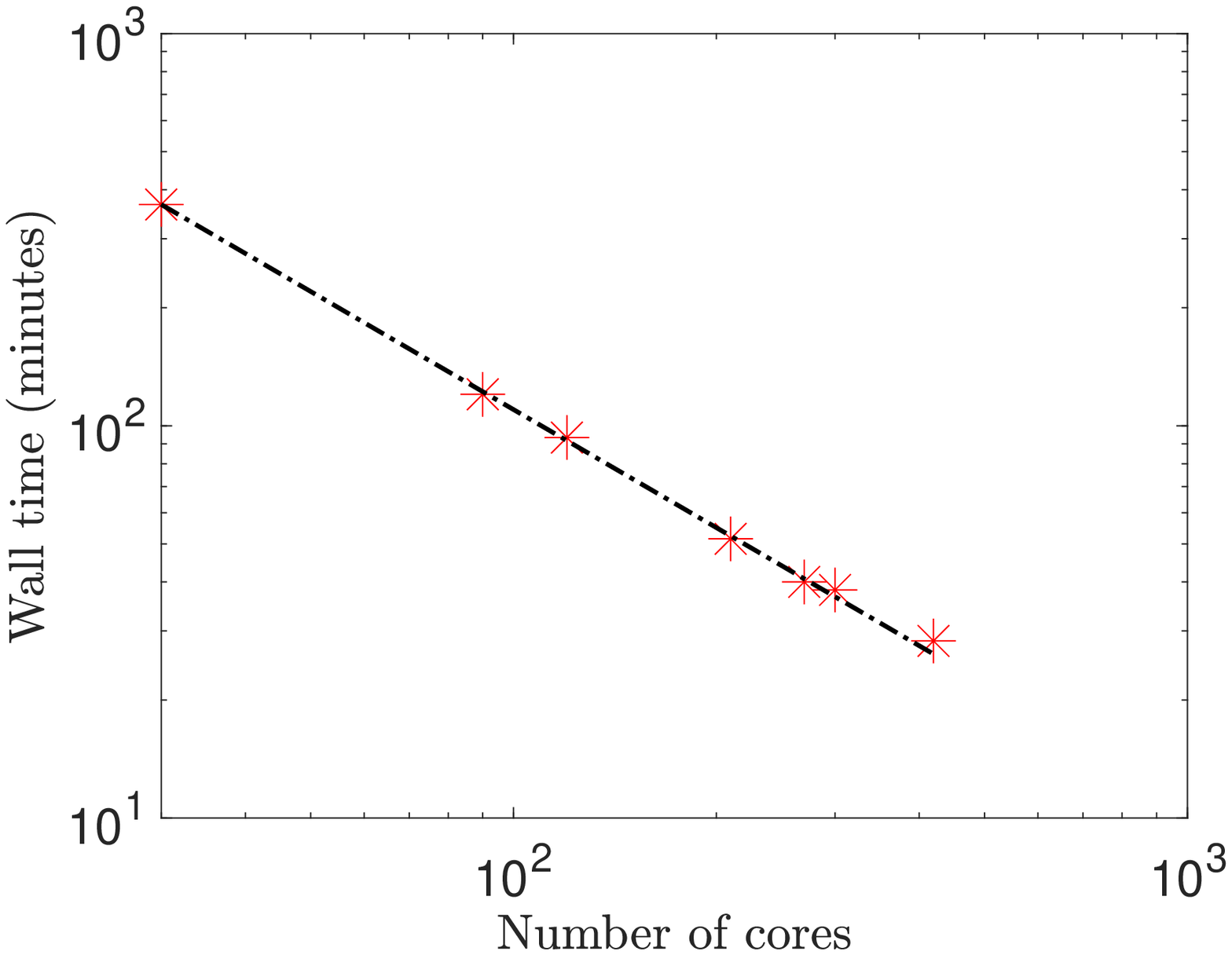}\label{Fig:StrongScaling}}
\subfloat[speed up]{\includegraphics[keepaspectratio=true,width=0.49\textwidth]{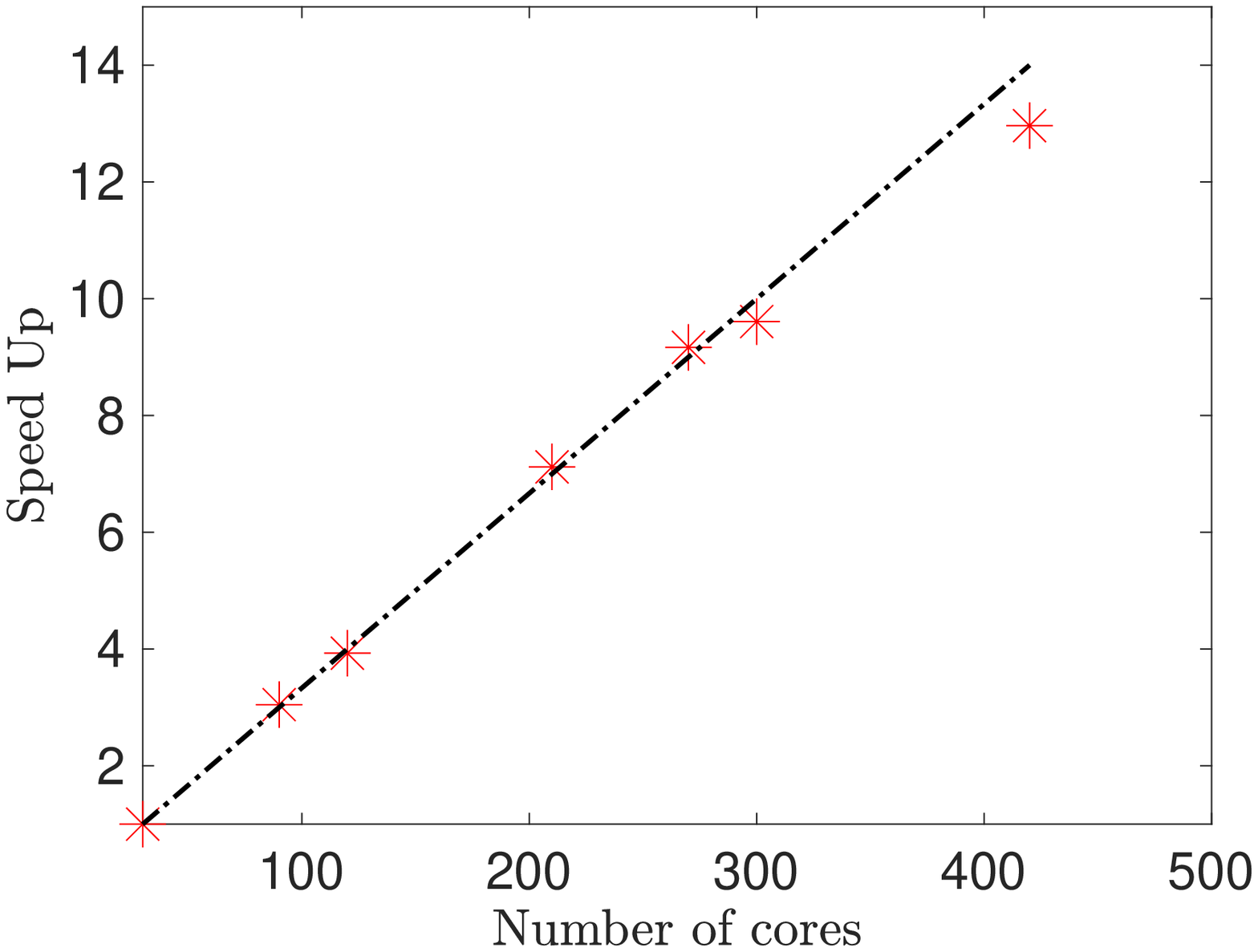}\label{Fig:SpeedUp}}\\
\subfloat[weak scaling]{\includegraphics[keepaspectratio=true,width=0.49\textwidth]{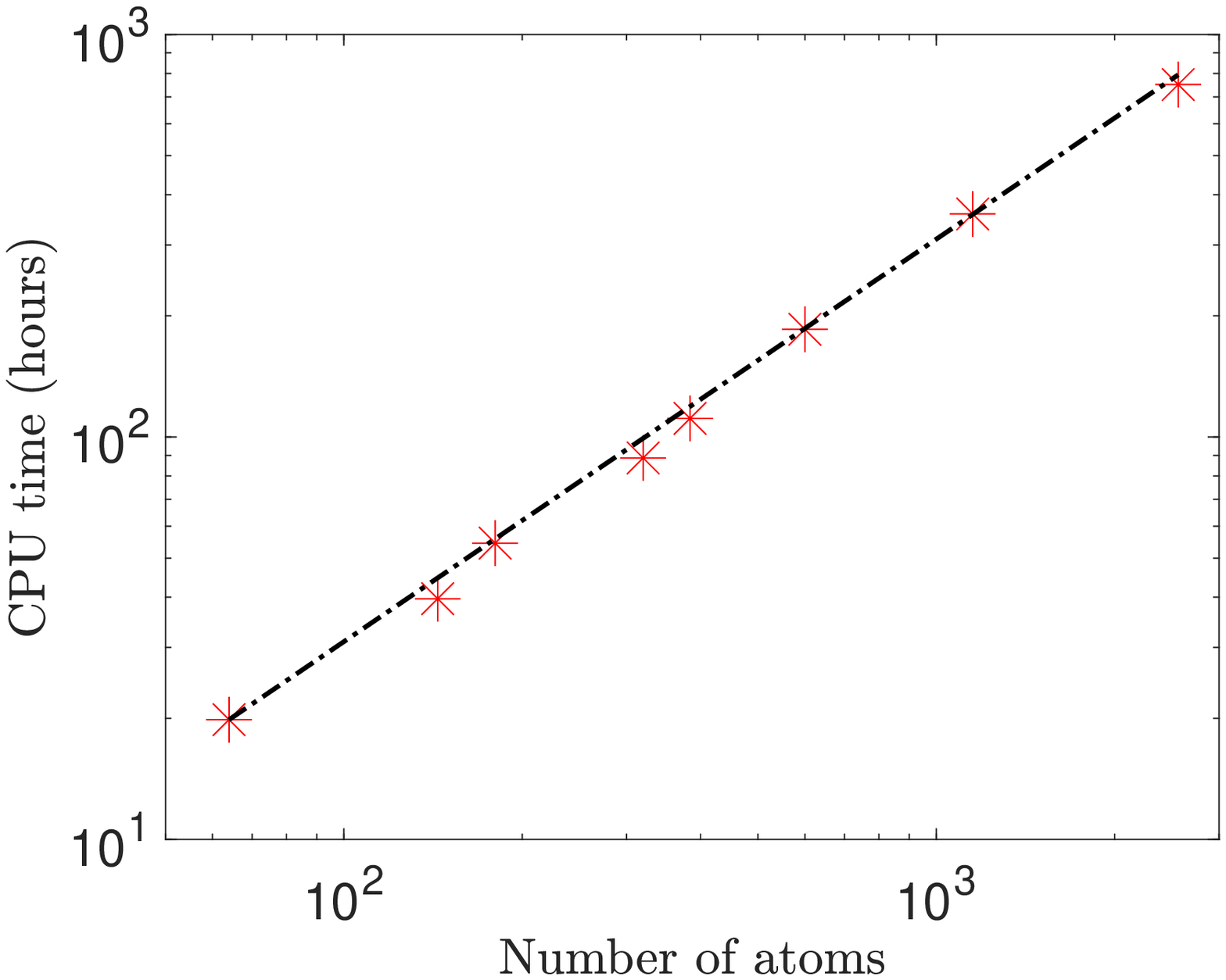}\label{Fig:WeakScaling}}
\subfloat[memory scaling]{\includegraphics[keepaspectratio=true,width=0.49\textwidth]{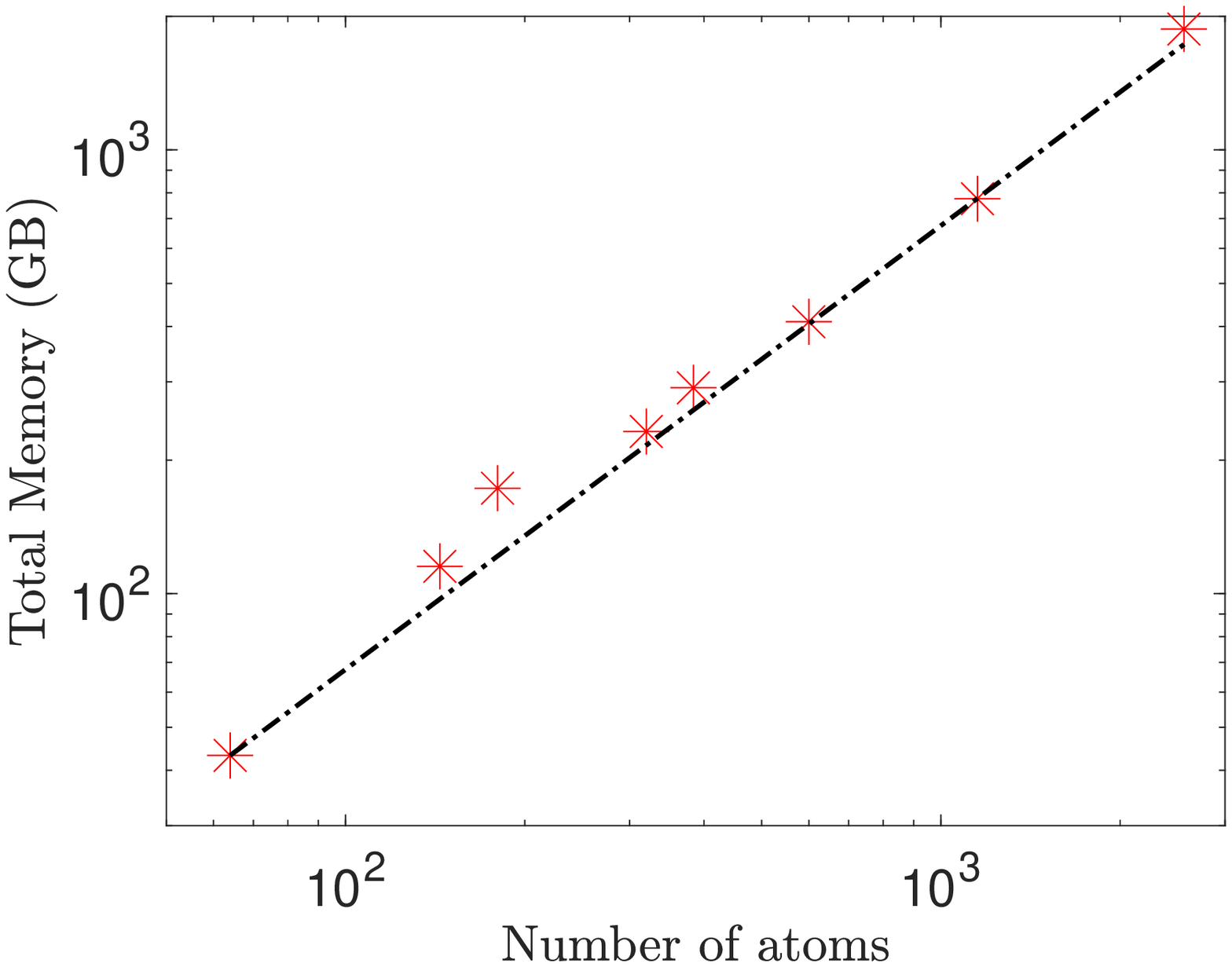}\label{Fig:MemoryScaling}}\\
\caption{Scaling and performance of the framework. The dash dot line is the ideal linear scaling behavior.} 
\end{figure}

Next, we turn to the scaling and efficiency of the formulation and parallel implementation developed in this work. We choose magnesium crystal with one aluminum solute atom, placed at the center of the computational domain. The simulation parameters used are $h$ = $0.6$ Bohr, $R_{cut}$ = $12.0$ Bohr, $K_L$ = $80$ and $K_C$ = $100$ 
with a desired accuracy of $2 \times 10^{-4}$ Ha/atom in energy and $2 \times 10^{-4}$ Ha/Bohr in atomic force.

We first perform a strong scaling study with a $600$ atom system, varying the number of cores from $30$ to $420$. The wall times for each SCF iteration is presented in Fig \ref{Fig:StrongScaling}. The parallel efficiency on $420$ relative to 30 cores is $92.6$ percent.   This data is plotted in terms of speed up in Fig. \ref{Fig:SpeedUp}.  Next, we perform a weak scaling study by selecting systems with sizes ranging from $64$ atoms to $2560$ atoms, while increasing the number of processors from $30$ to $900$. We choose these such that the number of atoms per MPI process is between two and three.    Fig \ref{Fig:WeakScaling} shows that the variation of CPU time for one SCF iteration versus the number of atoms is linear ($\approx \mathcal{O}(N)$).  Fig. \ref{Fig:MemoryScaling} shows that the memory required also scales linearly with respect to the number of atoms.  All of this shows excellent scaling of our algorithm.  This is due in part to restricting much of the parallel communication to the MPI processes that are neighbors, and keeping the global communication to a minimum.  

We note that the spectral quadrature step accounts for greater than $98$ percent of the total time in each SCF iteration. The prefactor of spectral quadrature can be significantly reduced by incorporating reduced basis methods such as Discrete Discontinious Basis Projection (\cite{Xu:2018}).  Further, the number of SCF iterations to achieve a fixed target SCF error increases with system size in metallic systems due to charge sloshing \cite{Kerker:1981}.  The introduction of real space preconditioning schemes \cite{Shiihara:2008,Kumar:2019} is likely to reduce this for large metallic systems.

\section{Defects in magnesium} \label{Section:Defects}
We now study isolated point defects and defect pairs in magnesium.  Of particular interest are the formation energy of isolated defects and binding energy of defect pairs.  
Let $\mathcal{E}(M,n,m)$ be the energy of a crystal with $M$ solvent atoms, $n$ solute atoms and $m$ vacancies. The formation energy of a defect cluster with $n$ solute atoms and $m$ vacancies is the excess energy of the crystal with defect cluster over the those of perfect crystals of the host and solute:
\begin{equation}\label{Eqn:Formation}
E_{n,m}^f = \mathcal{E}(M-n-m,n,m)- \frac{M-n-m}{M} \mathcal{E}(M,0,0) - n  \mathcal{\bar{E}},
\end{equation}
where $\mathcal{\bar{E}}$ is the energy per atom of the solute in its perfect crystalline state.   Note that when we have an isolated solute and no vacancies, the formation energy $E^f_{1,0}$  is referred to as  dilute impurity energy.  Further, the binding energy of this cluster is
\begin{equation}\label{Eqn:Binding}
E_{n,m}^b = n E_{1,0}^f + m E_{0,1}^f - E_{n,m}^f .
\end{equation}
Note that the defect is stable when the formation energy is negative, and the defect cluster has favorable binding when the binding energy is positive. 

\paragraph{Isolated point defect}
\begin{figure}[t!] 
\centering
\subfloat[Formation energy of a vacancy]{\includegraphics[keepaspectratio=true,width=0.49\textwidth]{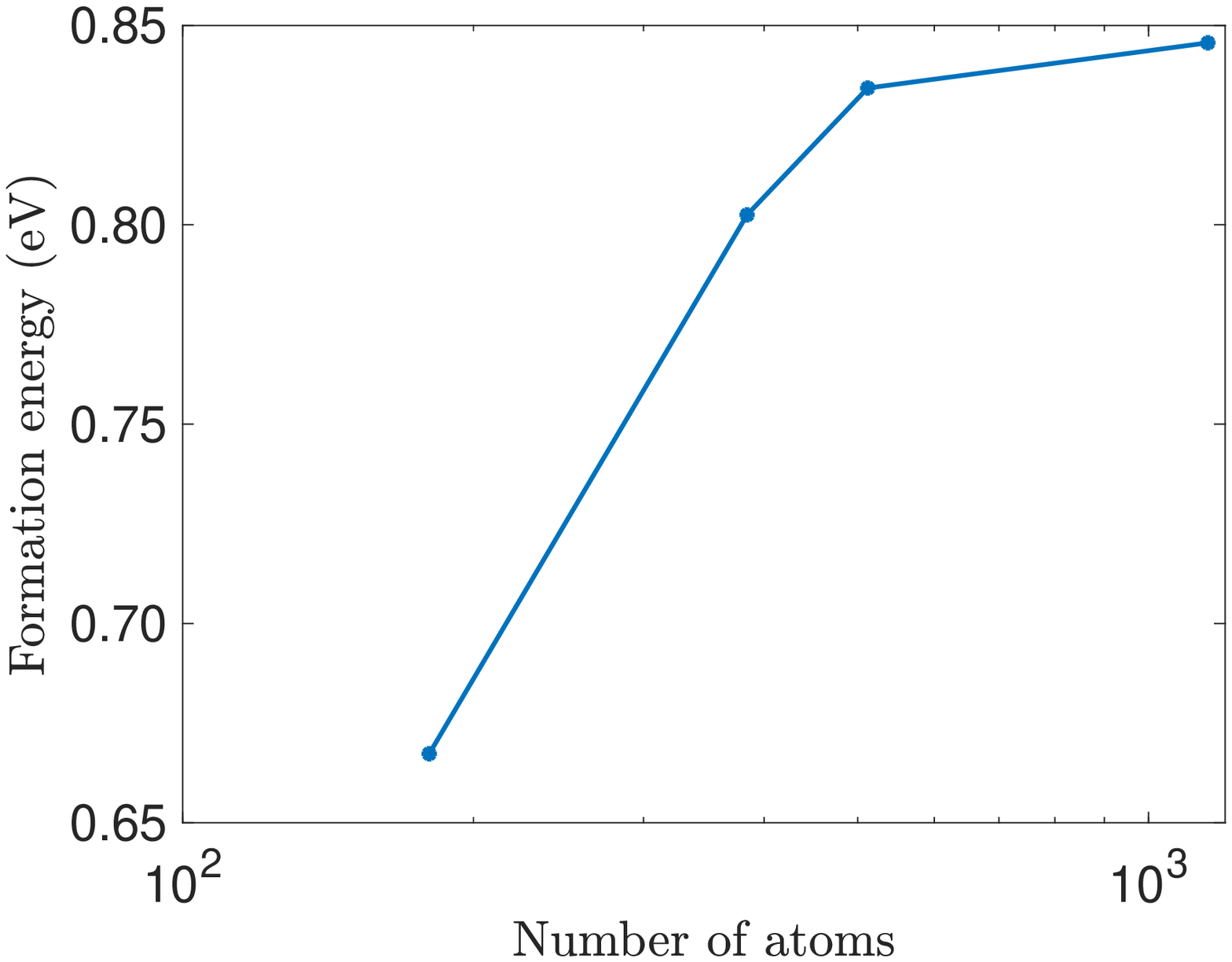}\label{Fig:VFENatoms}} 
\subfloat[Electron density near a vacancy]{\includegraphics[keepaspectratio=true,width=0.49\textwidth]{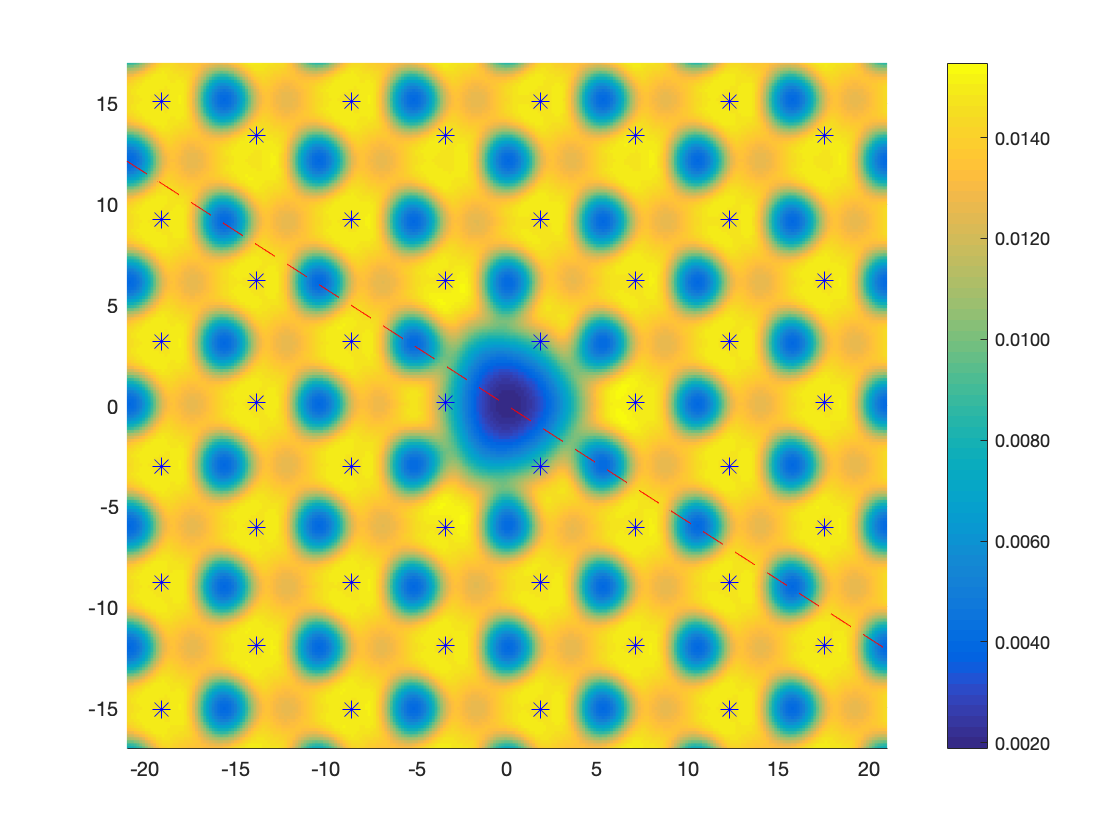}\label{Fig:Vacancy}}\\
\subfloat[Dilute impurity energy of Al]{\includegraphics[keepaspectratio=true,width=0.49\textwidth]{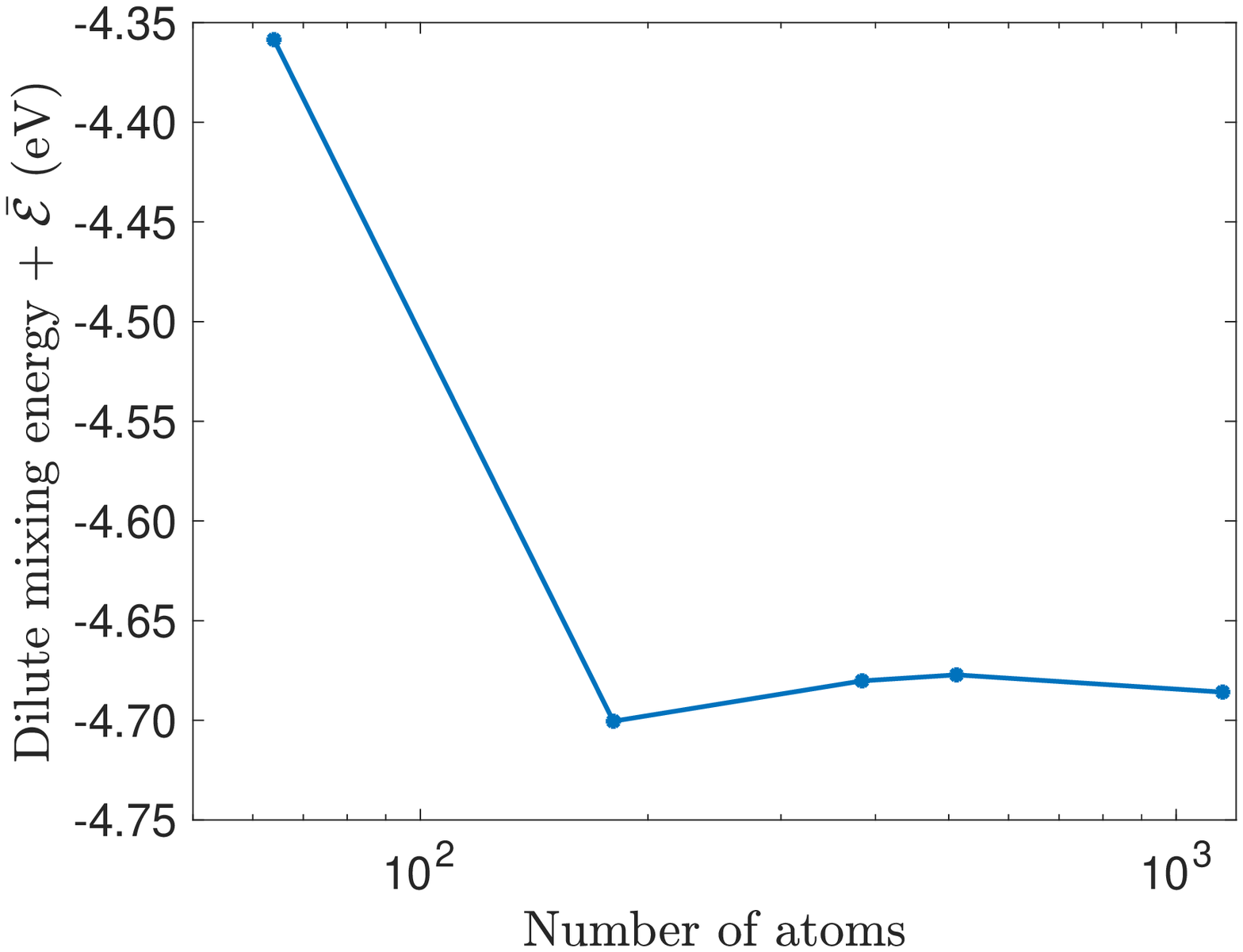}\label{Fig:SFENatoms}}
\subfloat[Electron density near a Al solute]{\includegraphics[keepaspectratio=true,width=0.49\textwidth]{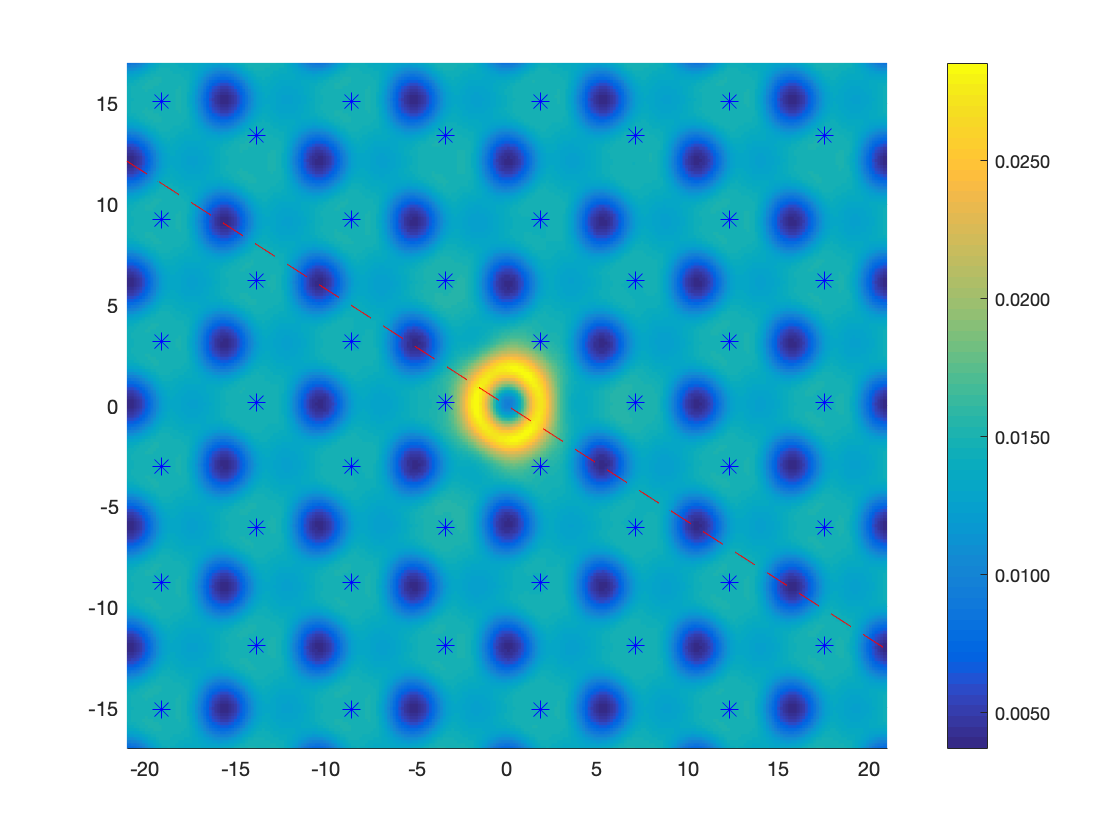}\label{Fig:Solute}}
\caption{Vacancy (a,b) and an aluminum solute (c,d) in magnesium.  The calculated vacancy formation energy (a) and dilute impurity energy (c) for various computational cell size and computed electron density along the basal plane for a vacancy (b) and solute (d).  The `*' marks in (b,d) indicate the projected positions of the atoms in the basal plane at a height $c/2$ above and below this plane.} 
\end{figure}

We calculate the formation energy of a monovacancy for various computational domain size from $63$ atoms to $1151$ atoms, and the results are shown in Fig. \ref{Fig:VFENatoms}.  We observe that the formation energy strongly depends on cell size, and converges at cell sizes of approximately $1000$ atoms to $0.8456$ eV.   This is broadly in agreement with values reported in the literature:  calculated values $0.779$ - $0.768$ eV using local pseudo-potential and a coarse grained approach with $1024$ to $1$ billion atoms \cite{Ponga:2016} and measured values of $0.58$ - $0.90$ eV \cite{Mairy:1967,Janot:1970,Tzanetakis:1976,Vehanen:1981}.
Fig. \ref{Fig:Vacancy} shows the electron density on the basal plane in the vicinity of the vacancy.  Unsurprisingly, it is depleted at the vacancy.  An interesting feature is that the electron density does not display the reflection symmetry of the basal plane (e.g. about the  red dashed line).  This is because the three dimensional crystal breaks this symmetry at the plane at a height $c/2$ above and below this plane.  This is emphasized by indicating the atoms on this upper and lower planes with a `*' in the figure.  This observation plays a role in binding.

Next we compute the dilute impurity energy of an aluminum solute atom, and this is shown in Fig. \ref{Fig:SFENatoms}.  We again observe that this energy depends on the cell size and converges at a few hundred atoms. The electron density on the basal plane is shown Fig. \ref{Fig:Solute}. The electron density if elevated near the Al solute and the distribution is more symmetric.

\paragraph{Defect pairs}

\begin{figure} \centering
\subfloat[Nearest neighbor positions]{\includegraphics[keepaspectratio=true,width=0.3\textwidth]{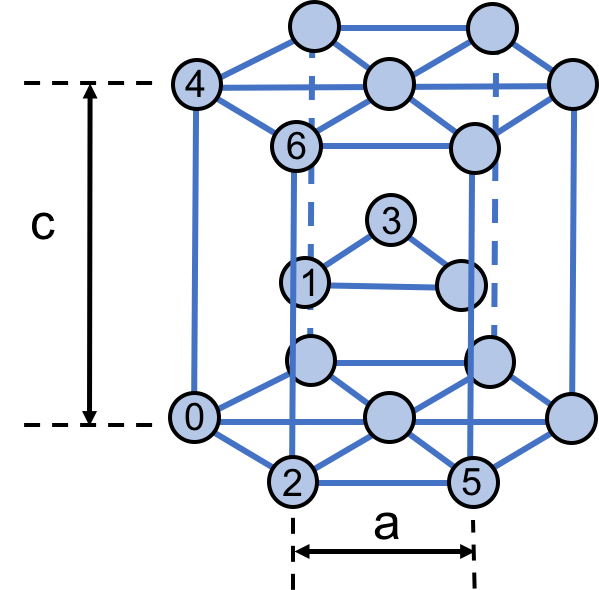}\label{Fig:NN}}
\subfloat[Vacancy-vacancy binding energy]{\hspace{1in}\includegraphics[keepaspectratio=true,width=0.49\textwidth]{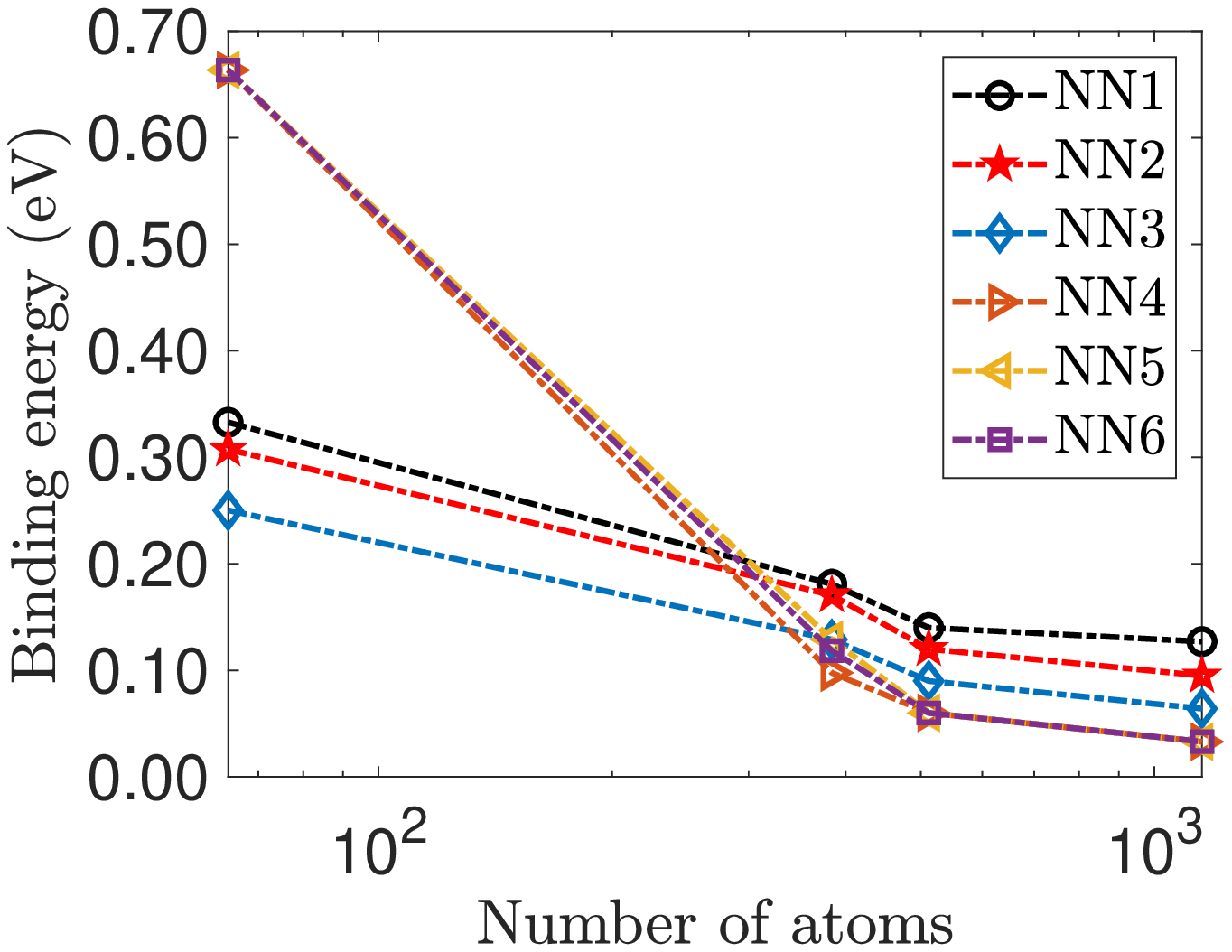} \label{Fig:VV} }\\
\subfloat[Vacancy-solute binding energy]{\includegraphics[keepaspectratio=true,width=0.49\textwidth]{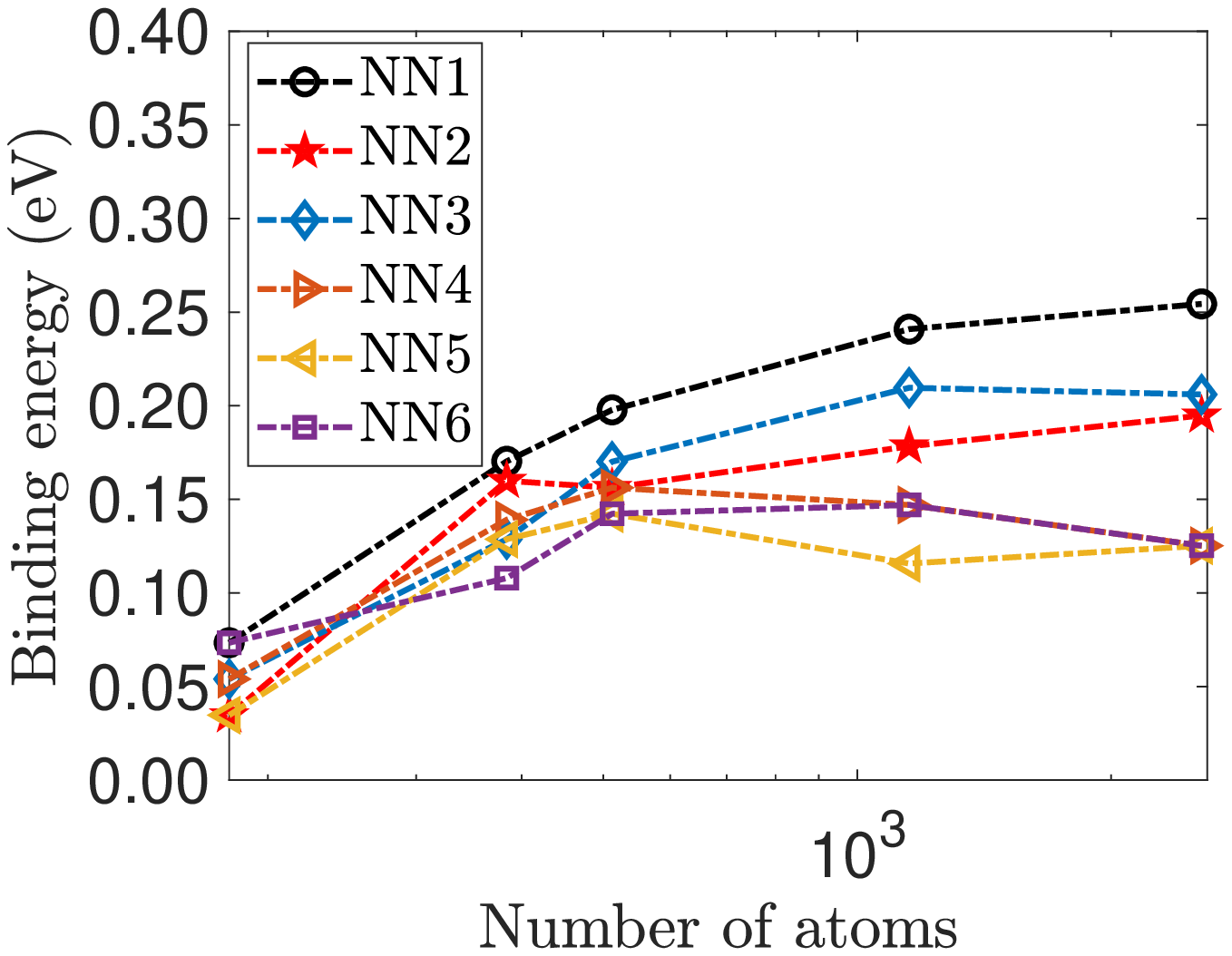}\label{Fig:VS}}
\subfloat[Solute-solute binding energy]{\includegraphics[keepaspectratio=true,width=0.49\textwidth]{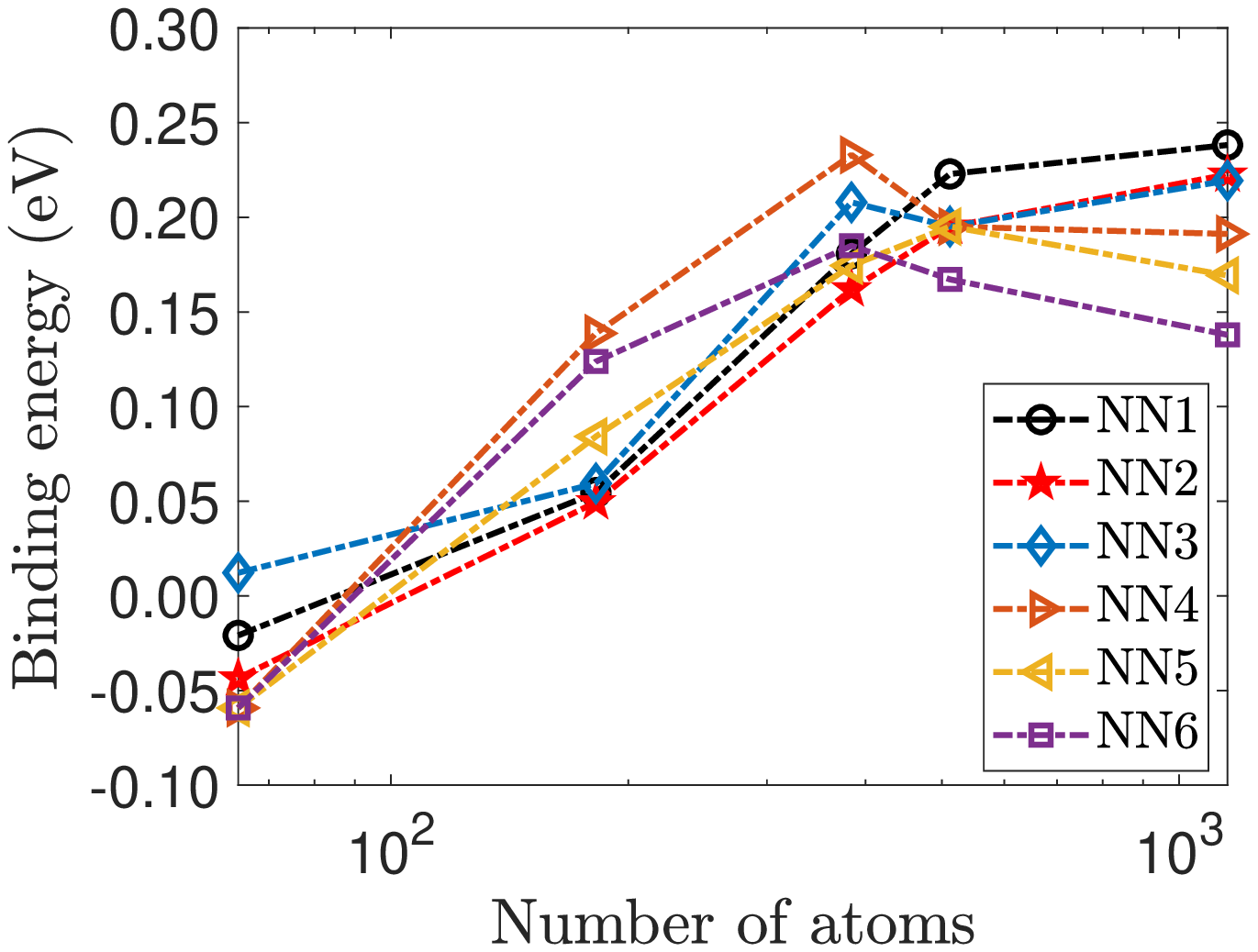}\label{Fig:SS}}
\caption{(a) Nearest neighbor positions for two point defects.  The first defect is at the point 0, and the second defect at one of the sites marked 1 through 6 in increasing order  distance. Computed (b) Vacancy-vacancy, (c) Vacancy-solute and (d) Solute-solute binding energies as a function of computational cell size. \label{Fig:binding}} 
\label{Fig:pairs}
\end{figure}

\begin{table}[t]
\caption{Defect pair binding energies.}
\centering
\begin{tabular}{ccccc}
\hline
NN & Experiment & Previous DFT (eV) & Previous DFT (eV) &  This work (eV)   \\
\hline
\multicolumn{5}{c}{Vacancy-vacancy}\\
1 &&  0.06$^a$ & 0.106$^b$ & 0.127  \\
2 &&  0.07$^a$ &  0.096$^b$ & 0.095 \\
3 &&  -0.01$^a$ & 0.057$^b$ & 0.064 \\
4 && 0.01$^a$ &  0.048$^b$ & 0.033 \\
5 && 0.01$^a$ &  0.038$^b$ & 0.033 \\
6 && 0.01$^a$ &  0.034$^b$ & 0.033 \\
\hline
\multicolumn{5}{c}{Vacancy-solute}\\
1 &  0.29 \cite{Rao:1978} &  0.06$^c$ && 0.2544 \\
2 & & & 0.03$^d$ & 0.1948 \\
3 &  & && 0.2060  \\
4 & & &&0.1252  \\
5 & & &&0.1252 \\
6 &  &  &&0.1252  \\
\hline
\multicolumn{5}{c}{Solute-solute}\\
1 &&& -0.021$^e$ &  0.2382 \\
2 && -0.21$^f$ & -0.018$^e$ & 0.2226 \\
3 && -0.24$^f$ & 0.024$^e$ & 0.2194 \\
4 &&& 0.014$^e$ & 0.1912 \\
5 &&&                       & 0.1693 \\
6 &&&                       &  0.1379 \\
\hline \hline
&\multicolumn{4}{l}{a. 96 atoms using Troullier-Martins pseudopotential  \cite{Uesugi:2003} }\\
&\multicolumn{4}{l}{b. 3456 atoms local pseudopotential \cite{Ponga:2016} }\\
&\multicolumn{4}{l}{c. 192 atoms using Projector Augmented Wave (PAW) pseudopotential \cite{Agarwal:2018}  }\\
&\multicolumn{4}{l}{d. 64 atoms using  ultrasoft pseudopotential  \cite{Shin:2010}  }\\
&\multicolumn{4}{l}{e. 96 atoms using Projector Augmented Wave (PAW) pseudopotential \cite{Kimizuka:2013} }\\
&\multicolumn{4}{l}{f. 64 atoms using Projector Augmented Wave (PAW) pseudopotential  \cite{Liu:2015} }\\
\hline
\end{tabular}
\label{Table:binding}
\end{table}

We now study the interaction between a pair of point defects: vacancy-vacancy, vacancy-solute and solute-solute.  Fig. \ref{Fig:NN}, shows the six possible nearest neighbor positions on a  HCP lattice where one defect is placed at the site marked 0 while the other defects are placed in the sites marked $1$ through $6$ denote the first six nearest neighbor configurations.   

Table \ref{Table:binding} shows the computed binding energy (cf. (\ref{Eqn:Binding})) for various defect pairs.  Recall that a positive binding energy indicates an energetic propensity for the defects to bind.  We observe that all three types of defect pairs -- vacancy pair, vacancy-solute and solute-pair -- have an energetic preference for binding.   This is especially strong for vacancy-solute pairs and solute pairs.   Further the binding energy increases with decreasing inter-defect distance in all three cases, indicating that the binding is stronger with decreasing distance.   These results are consistent with the observation that voids form readily in Mg, Al has a high diffusivity in Mg and that Al-rich precipitates from readily in Mg.   Our results agree quantitatively with the only case (first nearest-neighbor vacancy-solute-pair) where we are aware of experimental observation \cite{Rao:1978} -- Table \ref{Table:binding}.

The table also shows that while our results agree with those of Ponga, Bhattacharya and Ortiz \cite{Ponga:2016} in the case of vacany pairs, they do not agree with other prior DFT calculations.  Specifically, our results predict much stronger binding compared to other results (\cite{Uesugi:2003} for vacancy pairs, \cite{Agarwal:2018,Shin:2010} for vacancy-solute, and \cite{Kimizuka:2013,Liu:2015} for solute pairs).  In fact, previous results \cite{Kimizuka:2013,Liu:2015} for solute pairs also predict negative binding energies in some instances.

The previous calculations \cite{Uesugi:2003,Agarwal:2018,Shin:2010,Kimizuka:2013,Liu:2015} all used relatively small computatonal cells (64-192 atoms) compared to the current results (1151 atoms).   Fig. \ref{Fig:binding} shows the binding energy of various defect pairs computed for various sizes of the computational domain.  We see a significant dependence of the binding energy on the size of the computational cell with convergence only at around a 1000 atoms.  Further, there is no common trend.  For a solute pair, small cells predict a lack of binding while large cells predict binding.   The vacancy-vacancy binding energy is large for small computational domains, but decreases with domain size. For vacancy-solute, we see that the binding energy is small for small computational domains, but increases with domain size.  This tells us that the disturbance in the electronic and atomistic fields decay slowly.  Further, the previous studies use periodic boundary conditions which means that the defects interact if the unit cell is too small.  We have verified that a periodic calculations with a 96 atom unit cell (using SPARC\cite{Ghosh:2017b} which uses a similar finite difference formulation as the current work) reproduces the result of Uesugi {\it et al.} \cite{Uesugi:2003} who used the same pseudopotential.

We also note that Ponga, Bhattacharya and Ortiz \cite{Ponga:2016} used a local pseudopotential.  In other words, the role of the pseudopotential is less important in these defects compared to that of the slow decay.

\begin{figure} \centering
\subfloat[vacancy pair, basal plane]{\includegraphics[keepaspectratio=true,width=0.49\textwidth]{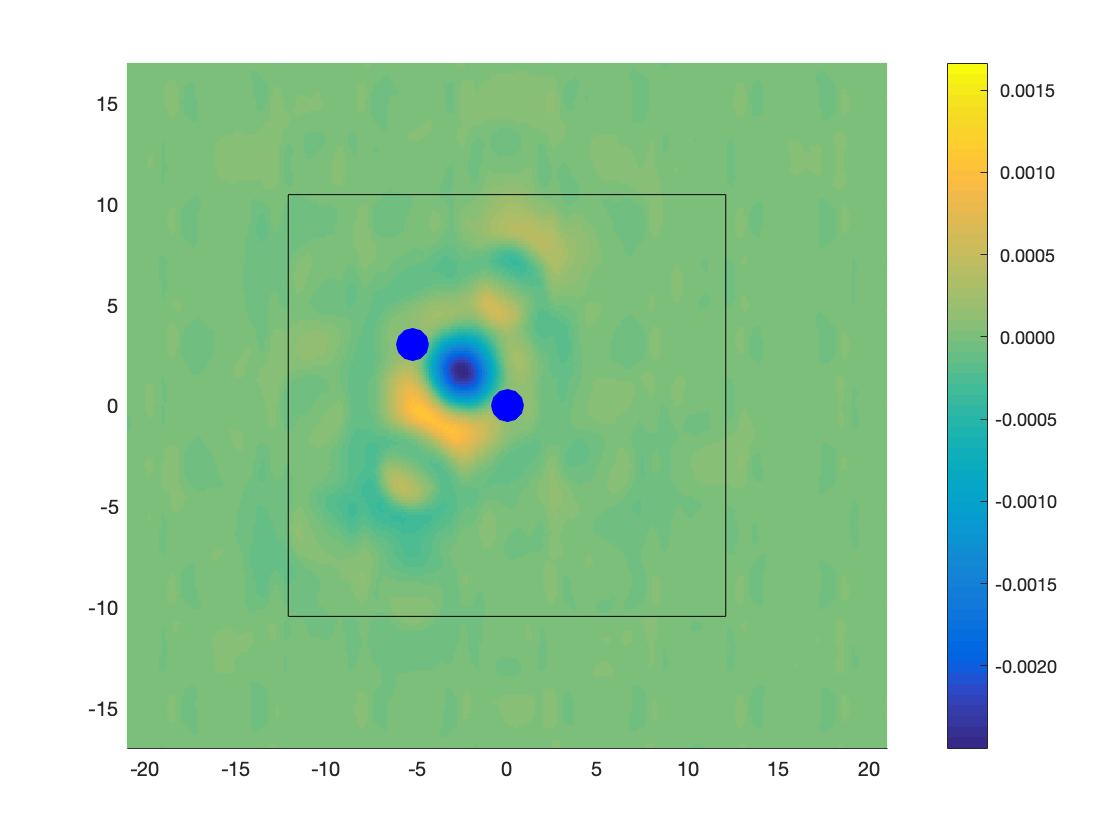}\label{Fig:VVNN2Basal}}
\subfloat[vacancy pair, prismatic plane]{\includegraphics[keepaspectratio=true,width=0.49\textwidth]{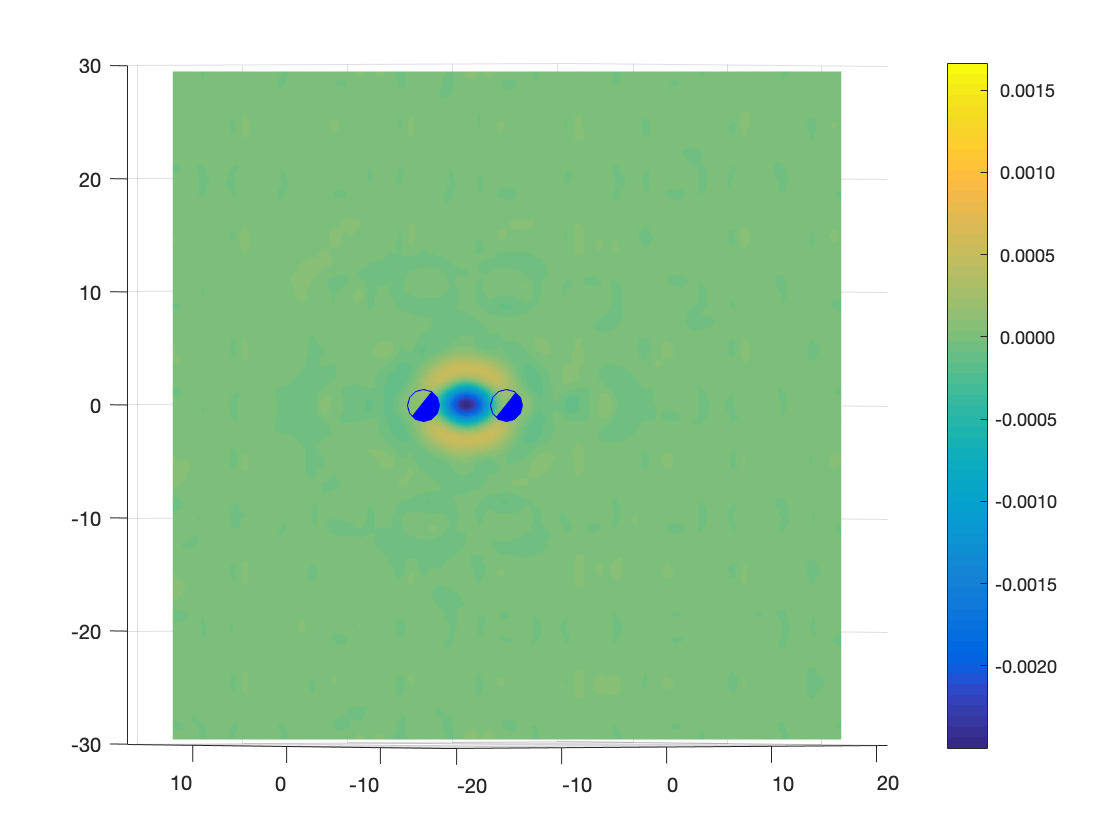}\label{Fig:VVNN2Prismatic}}\\
\subfloat[vacancy-solute, basal plane]{\includegraphics[keepaspectratio=true,width=0.49\textwidth]{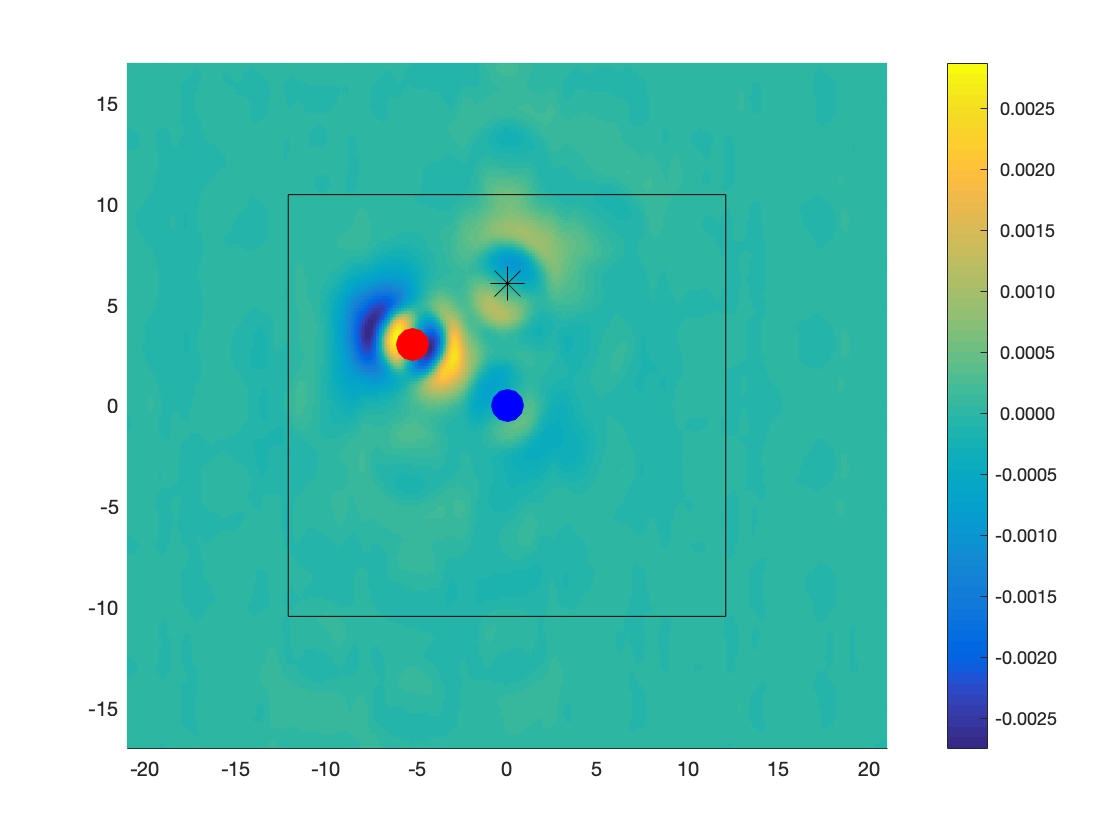}\label{Fig:SVNN2Basal}}
\subfloat[vacancy-solute, prismatic plane]{\includegraphics[keepaspectratio=true,width=0.49\textwidth]{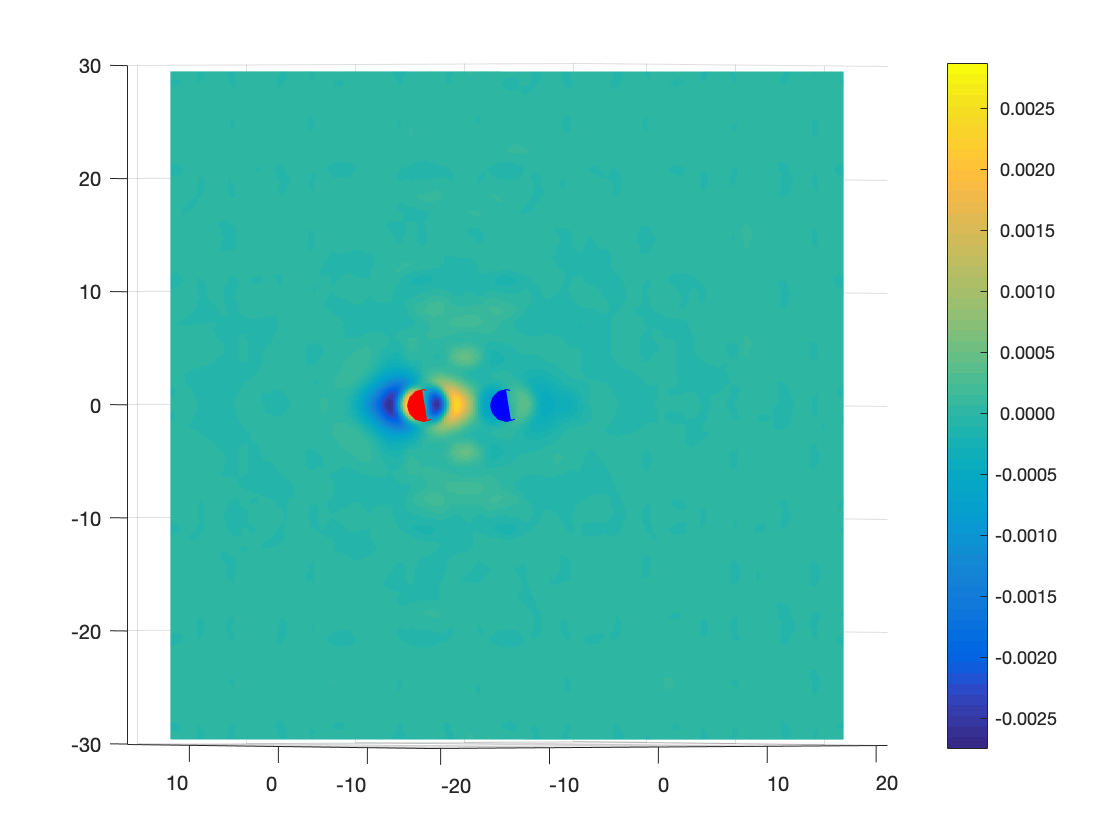}\label{Fig:SVNN2Prismatic}}\\
\subfloat[solute pair, basal plane]{\includegraphics[keepaspectratio=true,width=0.49\textwidth]{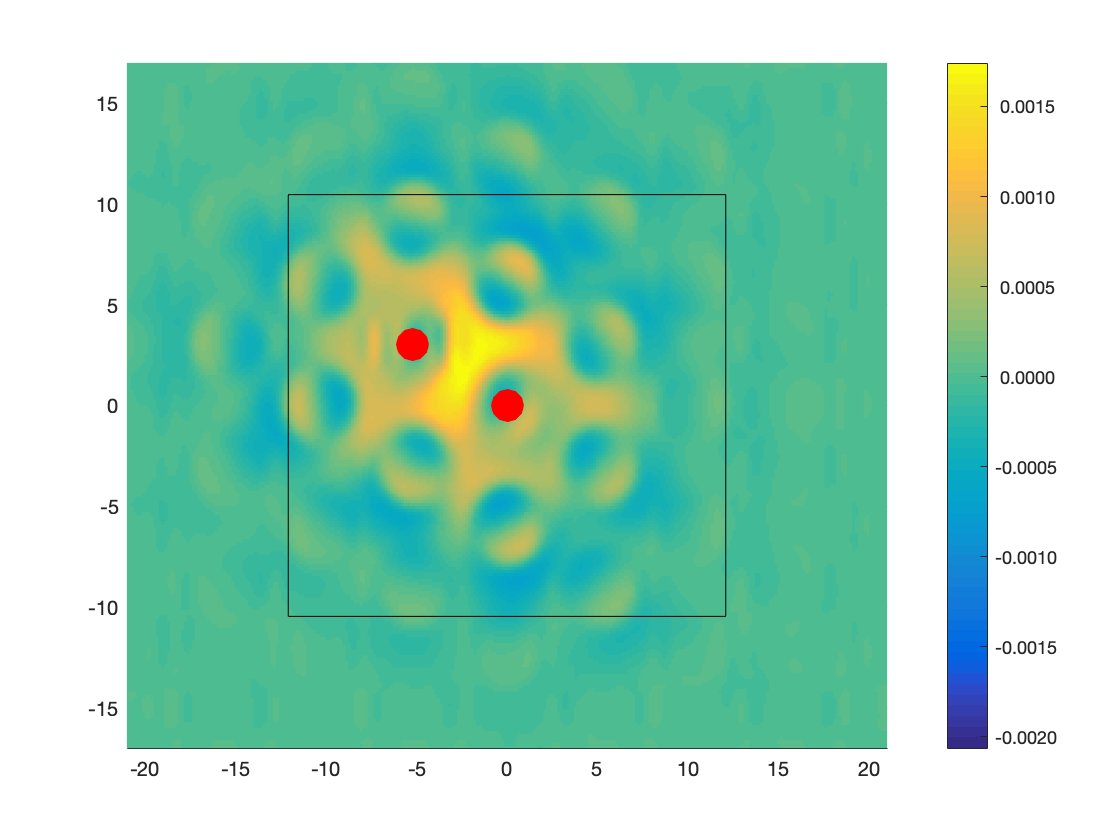}\label{Fig:SSNN2Basal}}
\subfloat[solute pair, prismatic plane]{\includegraphics[keepaspectratio=true,width=0.49\textwidth]{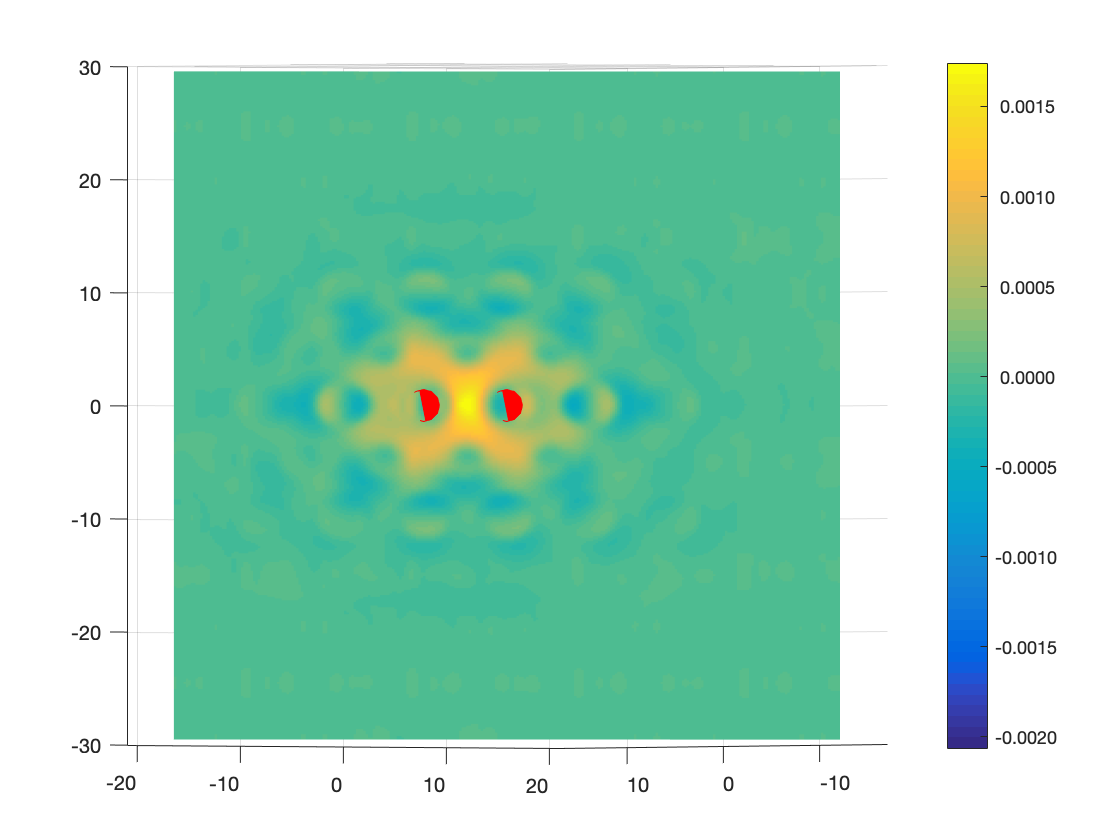}\label{Fig:SSNN2Prismatic}}
\caption{Electron density difference around a defect pair on the basal and prismatic plane.  The blue circles mark the position of the vacancies and the red circles that of the solutes.  The square box in (a,c,e) marks a 64 atom basal cell. \label{Fig:deltarho}}
\end{figure}

To understand the nature of the binding and explore further the decay, we examine the difference in electron density between that of a defect pair with that of a pair of defects:
\begin{equation}
\Delta \rho (\bx) = \rho^0(\bx)+\rho^n(\bx)-\rho_{p}(\bx)-\rho^{0n}(\bx) \,
\end{equation}
where $\rho^0$ is the electron density in a system with a defect placed at site 0, $\rho^n$ is that with a defect at site $n$, $\rho_{p}$ is that of the perfect crystal, and $\rho^{0n}$ is that of a defect pair (we need $\rho_{p}$ to make the system charge neutral).  These are plotted for second nearest-neighbor pair (i.e, a basal pair) in Fig. \ref{Fig:deltarho}.    We see that the core of the vacancy pair is severely depleted of electron density while the center of the solute pair has excess electron density.  The vacany-solute complex increases electronic density and depletes it at the distal region of the solute.  Further, we see that oscillations spread over several atomic spacings before decaying.  Furthermore, we see that the electron density in the case of vacancy and solute pairs are not symmetric with respect to the axis through the defects.  This is because of the influence of atoms on the basal plane just above and below the plane of observations.  Together, these mean that we need large computational cells for accurately computing the binding energies to accurately account for this decay.

\section{Concluding Remarks}\label{Section:Conclusion}
In this work, we have developed a computational framework that combines the computational efficiency of LSSGQ and enables the use of non-local pseudopotentials.   Further our implementation shows excellent strong and weak scaling.  This enables the application of DFT to problems that require a large computational domain such as defects in metals.  We have demonstrated the accuracy, efficiency and applicability of the framework by studying defects in magnesium.  In particular, we have shown that vacancy pairs, solute-vacancy pairs and solute pairs all bind strongly.   Importantly, the electron density decays slowly around these defects thereby necessitating large computational cells.

Our ongoing efforts include the development of an optimized implementation which is amenable to heterogeneous computer architectures. This is motivated from the observation that the calculation of spectral quadrature weights and nodes -- the computationally dominant part of the method -- is local to each process, and therefore can be accelerated on Graphics Processing Units (GPUs). The implementation of real-space preconditioning schemes \cite{Kumar:2019,Shiihara:2008} can further reduce the number of Self Consistent Field iterations for DFT simulations of large metallic systems. Finally, the incorporation of coarse-graining strategies \cite{Ponga:2016,Ponga:2020} will further reduce the scaling to sublinear with the number of atoms. Furthermore an accelerated calculation of the electronic predictor in the coarse grained framework will reduce the prefactor associated with scaling. Overall, these advances will render the first principles simulations of the interactions of extended defects such as dislocations tractable.

\section*{Acknowledgements}
This research was performed when S.G. held a position at the California Institute of Technology.   The work was sponsored by the Army Research Laboratory and was accomplished under Cooperative Agreement Number W911NF-12-2-0022. The views and conclusions contained in this document are those of the authors and should not be interpreted as representing the official policies, either expressed or implied, of the Army Research Laboratory or the U.S. Government. The U.S. Government is authorized to reproduce and distribute reprints for Government purposes notwithstanding any copyright notation herein. The computations presented here were conducted on the Caltech High Performance Cluster partially supported by a grant from the Gordon and Betty Moore Foundation, and the Extreme Science and Engineering Discovery Environment (XSEDE), which is supported by National Science Foundation grant number ACI-1548562. 

\bibliographystyle{abbrv}
\bibliography{LSDFTManuscript}

\end{document}